%% file: paper35_ML_MB.tex
\documentclass[twocolumn,english,pra]{revtex4-1}
\usepackage[T1]{fontenc}
\usepackage[latin9]{inputenc}
\usepackage{geometry}
\usepackage{mathrsfs}
\usepackage{amsmath}
\usepackage{amssymb}
\usepackage{graphicx}

\makeatletter
\usepackage{color}
\usepackage{braket}

\makeatother

\usepackage{babel}
\begin{document}

\title{Stroboscopic painting of optical potentials for atoms with subwavelength
resolution}

\author{M. Lacki$^{1}$, P. Zoller$^{2,3}$, M. A. Baranov$^{2,3}$}

\affiliation{\mbox{${}^1$Instytut Fizyki imienia Mariana Smoluchowskiego, Uniwersytet Jagiellonski, Lojasiewicza 11, 30-048 Krakow, Poland}
\mbox{${}^2$Center for Quantum Physics, Faculty of Mathematics, Computer Science and Physics, }
\mbox{University of Innsbruck, A-6020 Innsbruck, Austria}
\mbox{${}^3$Institute for Quantum Optics and Quantum Information of the Austrian Academy of Sciences, A-6020 Innsbruck, Austria}}
\begin{abstract}
We propose and discuss a method to engineer stroboscopically arbitrary
one-dimensional optical potentials with subwavelength resolution.
Our approach is based on subwavelength optical potential barriers
for atoms in the dark state in an optical $\Lambda$ system, which
we use as a stroboscopic drawing tool by controlling their amplitude
and position by changing the amplitude and the phase of the control
Rabi frequency in the $\Lambda$ system. We demonstrate the ability
of the method to engineer both smooth and comb-like periodic potentials
for atoms in the dark state, and establish the range of stroboscopic
frequencies when the quasienergies of the stroboscopic Floquet system
reproduce the band structure of the time-averaged potentials. In contrast
to usual stroboscopic engineering which becomes increasingly accurate
with increasing the stroboscopic frequency, the presence of the bright
states of the $\Lambda$-system results in the upper bound on the
frequency, above which the dynamics strongly mixes the dark and the
bright channels, and the description in terms of a time-averaged potential
fails. For frequencies below this bound, the lowest Bloch band of
quasienergies contains several avoided-crossing coming from the coupling
to high energy states, with widths decreasing with increasing stroboscopic
frequency. We analyze the influence of these avoided crossings on
the dynamics in the lowest band using Bloch oscillations as an example,
and establish the parameter regimes when the population transfer from
the lowest band into high bands is negligible. We also present protocols
for loading atoms into the lowest band of the painted potentials starting
from atoms in the lowest band of a standard optical lattice. 
\end{abstract}
\maketitle

\section{Introduction\label{sec:Introduction}}

Generating arbitrary landscapes of optical potentials for atoms with
optical subwavelength resolution is a key challenge in designing many-body
systems and applications based on cold atoms, see, for example, \cite{Bloch2008,Lewenstein2012}.
Realization of this goal provides an unprecedented control over atomic
systems, beyond the familiar far-off resonant laser traps and optical
lattices. A particular example we have in mind is creation of an optical
lattice, or superlattice with spacing much shorter than the optical
wavelength. One motiviation to create such short-spacing lattices
is significantly increased energy scales compared to achievable temperatures
in implementing Bose and Fermi Hubbard, and spin models, i.e. the
promise to prepare quantum phases which are not accessible in standard
setups. A second example, is to generate optical potentials with optical
barriers of arbitrary shape, compared to the sinusoidal potentials
generated by standing light waves. In the present paper we address
the problem of `painting' arbitrary optical potential landscapes with
subwavelength resolution. The present work builds on the ability to
generate $\delta$-function like potentials as non-adiabatic corrections
in atomic dark states of the $\Lambda$ systems \cite{Lacki2016,Jendrzejewski2016,Wang2017}.
These potential peaks are used as ''drawing pencils'' of the optical
landscape, which is obtained as a time averaged potential by moving
rapidly the spatial position and changing height of these barriers;
i.e. we require the stroboscopic painting of the potentials to be
fast relative to the time scale of atomic motion in the resulting
potential. The present scheme goes beyond previous proposals \cite{dubetsky2002lambda,ritt2006fourier,yi2008state,salger2009bloch,lundblad2014observations,Goldman2014,nascimbene2015dynamic,Bukov2015},
by allowing the generation of an arbitrary potential landscape (within
the scale of the optical wavelength).

The paper is organized as follows. In Section II we describe our stroboscopic
painting procedure and present possible protocols for painting smooth
and comb-like potentials. Section III contains the description of
our approach to numerical calculations of the quasienergies and eigenfunctions
within the Bloch-Floquet scheme, which we then apply to the analysis
of the atomic motion in the dark-state channel (Section IV) and in
the full three-channel problem (Section V). We also establish there
the conditions on the stroboscopic frequency for the appearence of
well-defined low energy Bloch bands in the dark channel, and discuss
their properties. In Section VI we address the dynamics of the Bloch
states in the time-averaged potential by considering Bloch oscillations.
Section VII contains the description of two protocols for loading
atoms into the lowest Bloch band of the painted potentials, followed
by concluding remarks in Section VIII.

\section{Painting potentials using nanoscale barriers}

\label{sec:painting}
\begin{figure}
\includegraphics[width=8.6cm]{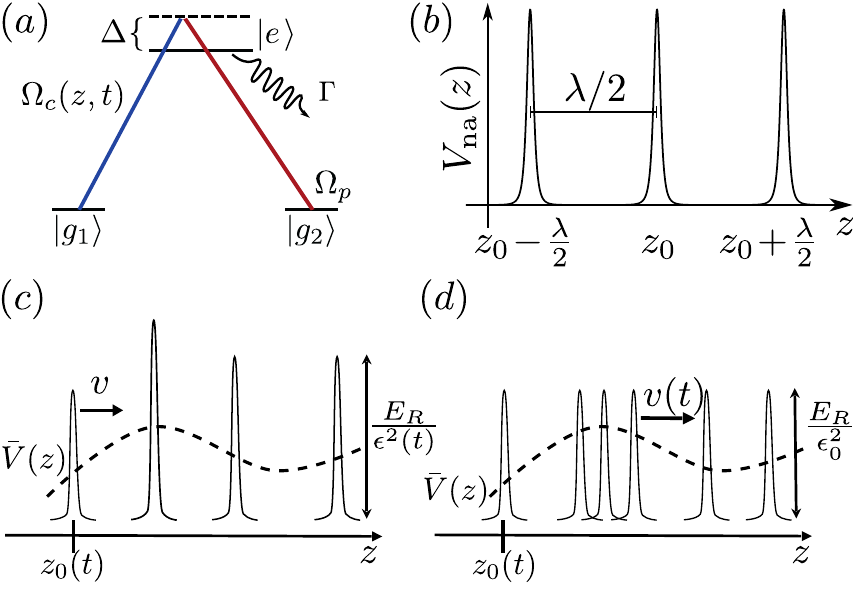}

\caption{Panel (a) shows the atomic optical $\Lambda$ scheme with two stable
atomic states $|g_{1}\rangle$, $|g_{2}\rangle$ and one excited state
$|e\rangle$, which is used for creation a subwavelength comb potential
shown in Panel (b) (see text). The Rabi frequency $\Omega_{p}$ is
time- and position-independent, while $\Omega_{c}(z,t)$ dependents
on both. Panels (c) and (d) show schematically two possible protocols
for stroboscopic painting of a potential $\bar{V}(z)$: In the Protocol
I in panel (c) peaks of the potential move with a constant velocity
and time-varying amplitude, while in the Protocol II in panel (d)
the potential peaks of a constant height move with varying velocity. }

\label{fig:Fig2atomicScheme}
\end{figure}
We start with a brief description of the setup for producing a periodic
comb of narrow $\delta$-function like optical potential peaks for
atoms using an optical $\Lambda$ system (for more details see \cite{Lacki2016}
and Appendix A). Let us consider an atom that can move along the $z$-direction,
while the motion in the other two transverse directions is frozen
to the ground state of a strong confining potential. We also assume
that the atomic internal structure allows formation of a well-isolated
optical $\Lambda$ system consisting of two stable internal states
$|g_{1}\rangle$ and $|g_{2}\rangle$ coupled to an excited state
$|e\rangle$ by the probe field and by the control one satisfying
the condition of the two-photon resonance, with the strengths given
by the Rabi frequencies $\Omega_{p}$ and $\Omega_{c}(z)$, respectively
{[}see Fig.~\ref{fig:Fig2atomicScheme} (a){]}, such that the Hamiltonian
takes the form
\begin{equation}
H(z)=-\frac{\hbar^{2}}{2m_{a}}\partial_{z}^{2}+H_{\Lambda}(z),\label{eq:hamiltonianqg1g2e}
\end{equation}
where
\begin{equation}
H_{\Lambda}(z)=\hbar\left(\begin{array}{ccc}
0 & \Omega_{c}(z)/2 & 0\\
\Omega_{c}(z)/2 & -\Delta-i\Gamma/2 & \Omega_{p}/2\\
0 & \Omega_{p}/2 & 0
\end{array}\right).\label{eq:threechannel}
\end{equation}
 The Rabi frequency $\Omega_{p}$ is $z$-independent, while $\Omega_{c}(z)$
has the form of a standing wave, 
\begin{equation}
\Omega_{c}(z)=\Omega_{c}\sin[k(z-z_{0})].\label{eq:standingwave}
\end{equation}
Because of the position dependence of $\Omega_{c}(z)$, the eigenstates
of the $\Lambda$ system in the adiabatic Born-Oppenheimer approximation
{[}i.e., the eigenstates of the Hamiltonian $H_{\Lambda}(z,t)${]}
are also position-dependent and include the dark state $|D(z)\rangle=-\cos\alpha(z)|g_{1}\rangle+\sin\alpha(z)|g_{2}\rangle$
with $\alpha(z)=\arctan[\Omega_{c}(z)/\Omega_{p}]$, which has zero
eigenenergy and is a linear combination of the stable atomic states
$|g_{1}\rangle$ and $|g_{2}\rangle$ only, and two bright states
$|B_{\pm}(z)\rangle$ containing also the excited state $|e\rangle$
and separated from the dark state by the gaps $\Delta E_{B\pm}=\min_{z}\left|\hbar E_{\pm}(z)\right|$,
respectively (see Appendix \ref{sec:details}). 

With the Hamiltonian of the kinetic energy {[}the first term in Eq.~(\ref{eq:hamiltonianqg1g2e}){]}
taken into account, an atom in the dark state $|D(z)\rangle$ experiences
a conservative nonadiabatic optical potential \cite{Lacki2016}, \cite{Wang2017} 

\begin{equation}
V_{\textrm{na}}(\epsilon,z-z_{0})=E_{R}\frac{\epsilon^{2}\cos^{2}[k(z-z_{0})]}{\{\epsilon^{2}+\sin^{2}[k(z-z_{0})]\}^{2}},\label{eq:Vsubwavelength}
\end{equation}
which depends on the ratio $\epsilon=\Omega_{p}/\Omega_{c}$ of the
amplitudes of the Rabi frequencies and the phase $kz_{0}$ of the
control field, $E_{R}=\hbar^{2}k^{2}/2m_{a}$ is the recoil energy,
$m_{a}$ is the mass of the atom. In addition, the kinetic energy
term generates nonadiabatic couplings between the dark and the bright
states. As shown in Refs.~\cite{Lacki2016} and {[}Experiment{]},
the effects of these couplings are small if the energy gaps $\Delta E_{B\pm}$
are much larger than the height $E_{R}/\epsilon^{2}$ of the potential
$V_{\textrm{na}}(\epsilon,z-z_{0})$, namely $\Delta E_{B\pm}\gg E_{R}/\epsilon^{2}$.
(For zero detuning $\Delta=0$, this condition requires $\Omega_{p}/2\gg E_{R}/\epsilon^{2}$.)
Under this condition, the motion of an atom in the dark state corresponds
to motion of a particle in the potential (\ref{eq:Vsubwavelength}).

\subsection{Stroboscopic painting}

For $\epsilon\ll1$, the potential $V_{\textrm{na}}(\epsilon,z-z_{0})$
has the form of a comb of sharp $\delta$-function like peaks centered
around $z_{0}+\pi n/k$ with integer $n$, which have a subwavelength
width $\sigma=\epsilon/k\ll\lambda$ (here $\lambda=2\pi/k$) and
a height $\hbar^{2}/(2m_{a}\sigma^{2})=E_{R}/\epsilon^{2}\gg E_{R}$
{[}see Fig.~\ref{fig:Fig2atomicScheme}(b){]}. These high and narrow
peaks can be used as ``pencils'' for drawing in the stroboscopic
way an arbitrary periodic potential with the spatial resolution $\sigma$
by fast changing the positions of the peaks and their height. This
can be achieved by taking a time-dependent control field 
\[
\Omega_{c}(z)\to\Omega_{c}(z,t)=\Omega_{c}(t)\sin\{k[z-z_{0}(t)]\}
\]
with the amplitude $\Omega_{c}(z,t)\gg\Omega_{p}$ which is periodic
in time with the period $T=2\pi/\omega$ and $z_{0}(t)$ periodic
or quasiperiodic, for example, $z_{0}(t+T)=z_{0}(t)+2\pi/k$. The
Hamiltonian (\ref{eq:threechannel}) and, hence, the total Hamiltonian
(\ref{eq:hamiltonianqg1g2e}) become now periodic functions of time
with the period $T$, $H_{\Lambda}(z,t)=H_{\Lambda}(z,t+T)$ and $H(z,t)=H(z,t+T)$.
{[}We note here that, although the spatial period of the potential
$V_{\textrm{na}}(\epsilon,z-z_{0})$ is $\pi/k$, the control field
$\Omega_{c}(z,t)$ and, therefore, the Hamiltonian have the spatial
period $2\pi/k$, and the stroboscopic approach has to be consistent
with the spatial periodicity of the Hamiltonian $H(z,t)$. Another
option, however, could be to use an anti-periodic function $\Omega_{c}(t)$,
$\Omega_{c}(t+T)=-\Omega_{c}(t)$, and $z_{0}(t+T)=z_{0}(t)+\pi/k$.{]}
The resulting nonadiabatic potential $V_{\textrm{na}}[\epsilon(t),z-z_{0}(t)]$
is then periodic in time with the frequency $\omega=2\pi/T$ and corresponds
to the moving comb of peaks with positions determined by $z_{0}(t)$,
and with the time-dependent amplitude $E_{R}/\epsilon(t)^{2}$.

When the stroboscopic frequency $\omega$ is much larger than the
typical frequency $\omega_{\mathrm{at}}$ of the motion of an atom
in the dark state but much smaller than the gaps $\Delta E_{B\pm}$
to the bright states, $\omega_{\mathrm{at}}\ll\omega\ll\Delta E_{B\pm}/\hbar$,
the atom in the dark state experiences the time-averaged potential

\begin{equation}
\bar{V}(z)=\frac{1}{T}\int_{0}^{T}V_{\textrm{na}}[\epsilon(t),z-z_{0}(t)]dt,\label{eq:runningapproximation}
\end{equation}
which is the leading term in the Magnus expansion \cite{goldman2014periodically}
based on the condition $\omega\gg\omega_{\mathrm{at}}$, and the effects
of the bright states can still be neglected. It is clear from the
previous discussion that, by choosing various functions $\epsilon(t)$
and $z_{0}(t)$, one can generate a large family of potentials $\bar{V}(z)$
restricted by only two constraints: \textsl{i}) The potential changes
on a spatial scale which is larger than $\sigma$ (or, in other words,
its spatial Fourier decomposition does not contain components with
the wave vector larger than $\sigma^{-1}$), and \textsl{ii}) It is
non-negative, $\bar{V}(z)\geq0$. The latter constraint does not actually
impose any physical limitations because one can always add a (positive)
constant to the potential of interest without changing the related
physics.

We note however that Eq.~(\ref{eq:runningapproximation}) for the
potential filled by an atom in the dark state, is valid only in the
ideal situation when $\omega_{\mathrm{at}}/\omega\to0$ and $\Delta E_{B\pm}/\omega\to\infty$.
For finite values of these ratios, Eq.~(\ref{eq:runningapproximation})
is only the leading approximation, and in the next sections we establish
the conditions which ensure the validity of this approximate solution
and find corrections to it. We address these issues on the two examples
of potentials: the one of the standard sinusoidal form and the other
of the comb-type, both with the shorter period $\lambda/2M$ ($M>1$is
an integer) as compared to the standard optical lattice and the comb
potential from Refs.~\cite{Lacki2016,Wang2017}.

\begin{figure}
\includegraphics[width=8.4cm]{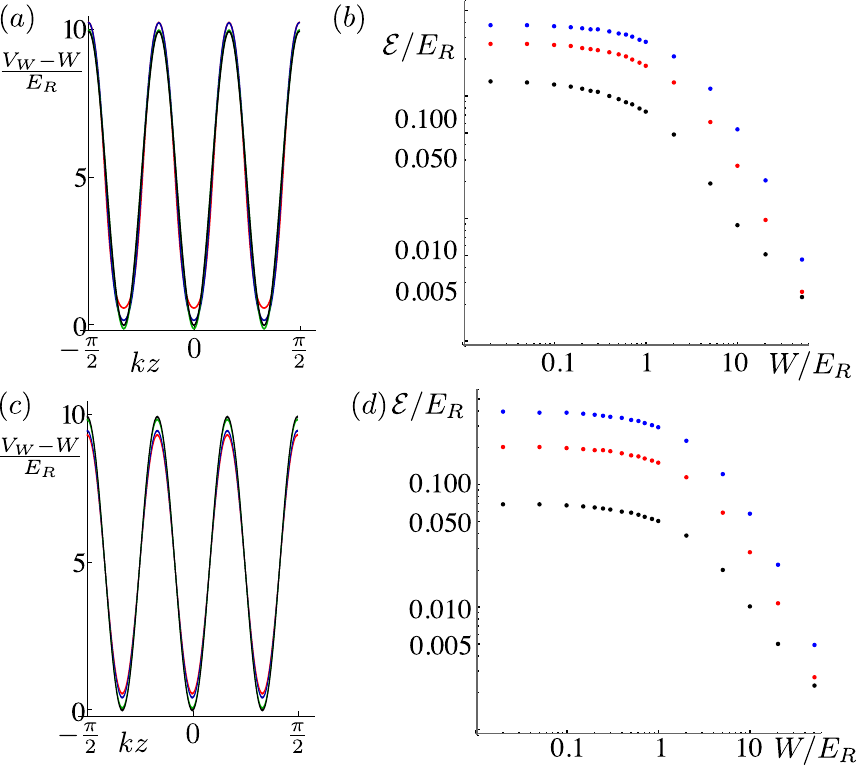}

\caption{(color online) Comparison of the two Protocols for painting being
applied to the potential $V(x)=V\sin^{2}(Mkx)+W$ with $V=10E_{R}$.
Panels (a) and (c) corresponds to the Protocol I, Eqs.~(\ref{eq:runningapproximation})
and (\ref{eq:epsequal}), and panels (b) and (d) to the Protocol II,
Eqs.~(\ref{eq:runningapproximation}) and (\ref{eq:epsequal2}).
Panels (a) and (c) show the painted potential $\overline{V}_{W}(z)$
for $M=3$ and $W=0.1,\,1,\,10E_{R}$ (red, blue, and green curves),
as well as the original potential $V(x)$ (black curve). The $W$-dependence
of the approximation error $||V-\overline{V}_{W}||_{L^{2}}$ for these
protocols is shown in panels (b) and (d), respectively, for $M=1,\,2,\,3$
(black, red, and blue dots, from bottom to top) (see text). }

\label{fig:AppendixApproximation}
\end{figure}

Following the above discussion, the stroboscopic ``painting'' of
an arbitrary optical potential $V(z)\geq0$ with spatial resolution
$\sigma$ starts with finding the functions $\epsilon(t)$ and $z_{0}(t)$
which solve Eq.~(\ref{eq:runningapproximation}) for $\bar{V}(z)=V(z).$
Note that there are (infinitely) many solutions of this problem, and
one can use this ambiguity to minimize the errors related to the approximate
character of (\ref{eq:runningapproximation}), including the effects
of the bright states. Below we discuss possible solutions (painting
protocols) of Eq.~(\ref{eq:runningapproximation}) for the cases
of a smooth potential (for precise definition, see below) and of a
comb-like one. It turns out that in the former case there exist two
simple approximate solutions of Eq.~(\ref{eq:runningapproximation})
for a general potential $V(z)$.

\subsection{Painting smooth potentials\label{sec:Painting-smooth-potentials}}

We consider first a smooth potential $V(z)$ that changes on a typical
scale $L$ such that $\lambda/2>L\gg\sigma$ and satisfies the condition
$V_{\mathrm{min}}>E_{R}$ with $V_{\mathrm{min}}$ being the minimal
value of $V(z)$, which, as discussed above, is not restrictive because
we can always add a constant to the potential. This condition originates
from the fact that the potential (\ref{eq:runningapproximation})
is strictly positive and its typical value $\bar{V}$ can be estimated
as a product of the typical height of the potential peak $V_{\mathrm{na}}\sim\hbar^{2}/(m_{a}\sigma^{2})$
and the typical fraction $\sim\epsilon$ of the period during which
the peak crosses a spatial region of the size $\sigma$ having a typical
velocity $v\sim\lambda/T$. We thus obtain $\bar{V}\sim E_{R}/\epsilon\geq E_{R}$.
Note also that in experiments, the width of the peaks $\sigma$ is
always limited, $\sigma\geq\sigma_{\mathrm{min}}=\epsilon_{\mathrm{min}}/k$,
by the available laser power and by the necessity of having as large
as possible gap between the dark and the bright states. This provides
an upper limit $V_{\mathrm{max}}\leq E_{R}/\epsilon_{\mathrm{min}}$
on the maximal value $V_{\mathrm{max}}$ of the smooth potential $V(z)$
which can be painted stroboscopically.

For a potential satisfying the above conditions, the functions $\epsilon(t)$
and $z_{0}(t)$ change slowly on the time scale $\sigma/v\sim\sigma T/\lambda$,
and can be easily found from Eq.~(\ref{eq:runningapproximation})
in two limiting cases, when either peaks of varying heights move with
a constant velocity (Protocol I) {[}see Fig.~\ref{fig:Fig2atomicScheme}(c){]},
or they have a fixed height but move with varying velocity (Protocol
II) {[}see Fig.~\ref{fig:Fig2atomicScheme}(d){]}. 

Let us first present the solution of Eq.~(\ref{eq:runningapproximation})
for the Protocol I when the peaks of the potential Eq.~(\ref{eq:Vsubwavelength})
move with a constant velocity $v=\lambda/T=\omega/k$, i.e. $z_{0}(t)=vt$.
It is easy to see that the desired function $\epsilon(t)$ {[}and,
therefore, $\Omega_{c}(t)${]} changes slowly on the time scale $\sigma/v$,
and that the main contribution to the averaged potential at position
$z$ comes from the value of $\epsilon(t)$ at time $t_{z}\approx z/v$.
We therefore obtain
\begin{eqnarray}
\epsilon(t) & \approx & \frac{E_{R}}{2V(vt)}\ll1,\label{eq:epsequal}
\end{eqnarray}
where we used the condition $V(z)\gg E_{R}$. Note that the spatial
resolution $\sigma=\epsilon/k$ in this Protocol is not homogeneous:
It is worse in the region of small values of $V(z)$. 

In the Protocol II with a constant height of the peaks, we have $\epsilon(t)=\epsilon_{0}\ll1$,
and the peaks, keeping their form, move with the varying velocity
$v=dz_{0}/dt$. The value of $\epsilon_{0}$ is related to the average
(over the spatial period $\lambda/2$) value $\overline{V}$ of the
potential $V(z)$ as 
\begin{equation}
\overline{V}\equiv(2/\lambda)\int_{0}^{\lambda/2}V(z)dz=E_{R}/(2\epsilon_{0}),\label{eq:eps0val}
\end{equation}
which follows from the fact that $(2/\lambda)\int_{0}^{\lambda/2}V_{\textrm{na}}[\epsilon_{0},z-z_{0}(t)]\textrm{d}z\approx E_{R}/(2\epsilon_{0})$.
The velocity $v(z)$ at which the peak should cross the point $z$
is

\begin{eqnarray}
v(z) & \approx & \frac{\pi E_{R}}{2\epsilon_{0}V(z)}\frac{1}{kT}=v_{0}\frac{\overline{V}}{V(z)},\label{eq:epsequal2}
\end{eqnarray}
where $v_{0}=\lambda/T$ is the average velocity, and the required
$z_{0}(t)$ can be found from the equation
\[
\frac{2}{\lambda}\int_{0}^{z_{0}(t)}\frac{V(z)}{\overline{V}}dz=\frac{t}{T}.
\]

\begin{figure}
\includegraphics[width=8cm]{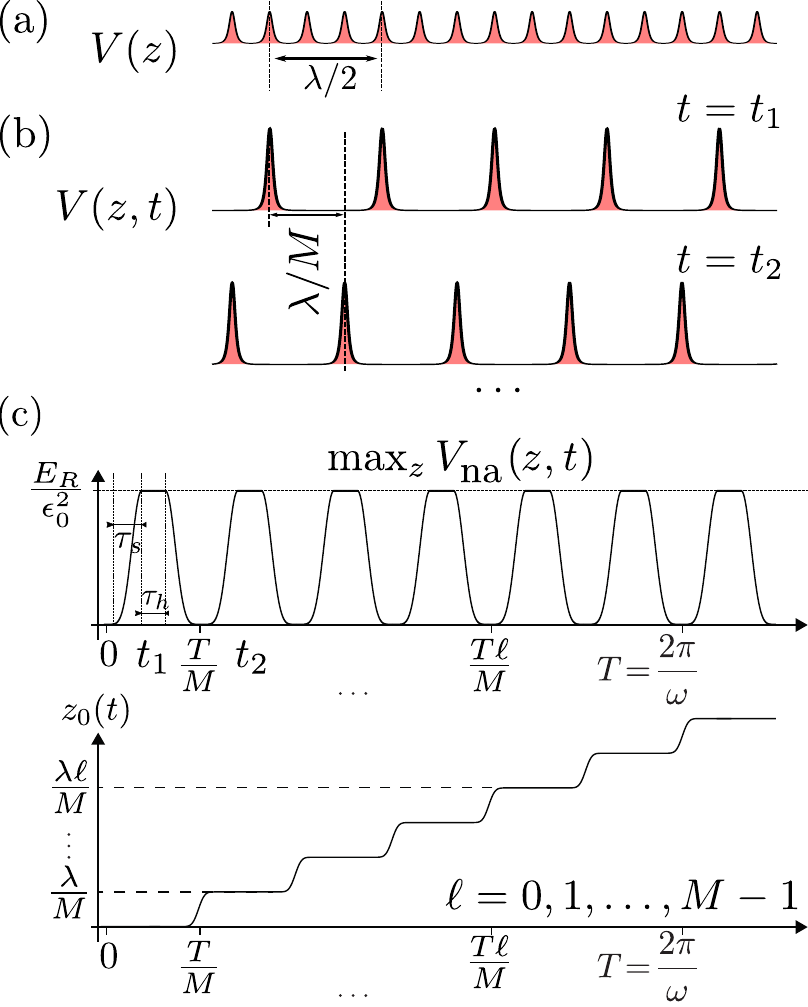}

\caption{(color online) A comb potential with $M=3$ peaks per length $\lambda/2$
{[}panel $(a)${]} and a ``flashing'' steps for its painting (Protocol
III) in panel (b), see text. Panel (c) shows functions $E_{R}/\epsilon^{2}(t)$
and $\bar{z}(t)$ used for this protocol.}

\label{fig:FigStroboscopic}
\end{figure}

As an illustration, let us consider the potential 
\begin{equation}
V_{W}(z)=V\sin^{2}(Mkz)+W,\label{eq:sinpotential}
\end{equation}
where $M$ is an integer ($M\geq1$) and $W$ is a positive offset,
which is an analog of a standard optical potential but with a shorter
lattice spacing $a=\lambda/(2M)$. The results of applying both protocols
with the approximate solutions from Eqs.~(\ref{eq:epsequal}) and
(\ref{eq:epsequal2}) are presented in Fig.~\ref{fig:AppendixApproximation}.
The left panels show the resulting painted potential $\overline{V}_{W}(z)$
calculated for $V=10E_{R}$ and three different values of the offset
$W/E_{R}=0.1$, $1$, and $10$, together with the original potential
$V_{W}(z)$. We see that in the Protocol I, following Eq.~(\ref{eq:epsequal}),
the subwavelength peak is smaller and broader near the minima of $V_{W}$
resulting in a larger error as compared to that in the maxima. In
the Protocol II, the height and the width of the peaks are constant
and the errors at the minima and maxima of $V_{W}$ are the same.
Panels (b) and (d) illustrate the mean-square error between the painted
potential $\overline{V}_{W}(z)$ and the original one $V_{W}(z)$,
\[
\mathcal{E}=||V_{W}-\overline{V}_{W}||_{L^{2}}=\{\int_{-\pi/2kM}^{\pi/2kM}[V_{W}(z)-\overline{V}_{W}(z)]^{2}dz\}^{1/2}.
\]
We see that generically the error decreases with increasing the potential
offset $W$ because it results in a smaller values of $\epsilon$
and, as a consequence, in a better spatial resolution $\sigma$. We
note, however, that for a smooth potential, the values $W\lesssim E_{R}$
suffice to obtain a very good results, see Fig.~\ref{fig:AppendixApproximation}.
On the other hand, the error increases with increasing $M$ (for a
fixed $W$), because the scale $L$ on which the potential changes
decreases with increasing $M$, and, having the same spatial resolution
$\sigma,$ one naturally expects a larger error in painting. To increase
the accuracy, therefore, one has to consider as high value of the
offset $W$ as possible. Another option is to optimize the time dependencies
of $\epsilon(t)$ and $z_{0}(t)$.

\subsection{Painting comb potentials\label{sec:Painting-comb-potential}}

Now we turn to the case of potentials with $L\sim\sigma$, when the
functions $\epsilon(t)$ and $z_{0}(t)$ vary on a short time scale
$\sigma T/\lambda$. Unfortunately, with a potential $V(z)$ of this
type, Eq.~(\ref{eq:runningapproximation}) for $\epsilon(t)$ and
$z_{0}(t)$ cannot be solved explicitly, except for special cases.
One of them is a periodic comb potential made of peaks with a width
$\sim\sigma$ separated by a distance $a=\lambda/(2M)$, where $M>1$
is an integer, which we consider below, see Fig. 3(a) for $M=3$.
This potential is an analog of a comb potential from Refs.~\cite{Lacki2016,Wang2017}
but with a shorter lattice spacing. A possible solution is given by
the Protocol III which corresponds to a sequence of stroboscopic ``flashes''
of the potential $V_{na}$ during the stroboscopic period $T$, with
some spatial shift of $V_{\mathrm{na}}$ from flash to flash, see
Fig. 3(b). Due to fact that the potential $V_{\mathrm{na}}(z)$ has
twice shorter period than the original Hamiltonian, a specific realization
of this protocol depends on the parity of $M$. For an odd $M$, it
is sufficient to have $M$ flashes of $V_{\mathrm{na}}$ per stroboscopic
period, and from flash to flash the potential $V_{\mathrm{na}}$ should
be shifted by the distance $2a=\lambda/M$. The details of this procedure
are presented in Fig. 3(c). The upper panel shows the time dependence
of the height of the potential $V_{\mathrm{na}}$: It first increases
from $0$ (this formally corresponds to $\epsilon=\infty$, or $\Omega_{c}=0$)
to its maximal value $E_{R}/\epsilon_{0}^{2}$ during time $\tau_{s}$,
then remains constant for the time $\tau_{h}$, and finally decreases
back to $0$ during time $\tau_{s}$. When the potential $V_{\mathrm{na}}$
stays close to zero, the function $z_{0}(t)$ rapidly increases its
value by the amount $\lambda/M=2a$, see lower panel in Fig.\ 3(c),
and then the process repeats. For an even $M$, one needs $2M$ flashes
of $V_{\mathrm{na}}$ per stroboscopic period, and from flash to flash
the potential $V_{\mathrm{na}}$ should be shifted by the distance
$a=\lambda/(2M)$.

Assuming the potential $V_{\mathrm{na}}$ being switched on/off linearly
in time $\tau_{s}$, we obtain from Eq.~(\ref{eq:runningapproximation})
that the generated potential $\bar{V}$ consists of equidistant peaks
of the height 
\begin{equation}
\max\bar{V}(z)=\frac{1}{M}\frac{E_{R}}{\text{\ensuremath{\epsilon}}_{0}^{2}}\alpha,\quad\alpha=\frac{\tau_{h}+\tau_{s}}{\tau_{h}+2\tau_{s}}<1,\label{eq:barV}
\end{equation}
and the width $\epsilon_{0}\lambda/2\pi$, separated by the distance
$a=\lambda/2M$, see Fig. 3(a). Being compared to the potential $V_{\mathrm{na}}(z)$,
the resulting potential $\bar{V}(z)$ has $M$ times shorter spatial
period, the potential peaks in $\bar{V}(z)$ have the same width as
in $V_{\mathrm{na}}(z)$, but their height is $M$ times smaller (
for $\tau_{s}\to0$). The above conclusions are practically independent
on how one switches on/off the potential during the time $\tau_{s}$;
a different choice results in small variations of the form of $\bar{V}(z)$
and the value of $\max\bar{V}(z)$. In the next Section we present
a detailed quantum mechanical Floquet analysis of the above schemes,
illustrated by examples both smooth and comb potentials with three
times short spatial period than that of $V_{\mathrm{na}}$ ($M=3$). 

\section{Floquet analysis and band structure }

\label{sec:Sec3}

In this section we perform a complete Floquet analysis of the time-periodic
three-level Hamiltonian$H(z,t)$ {[}see (\ref{eq:hamiltonianqg1g2e}){]}
to manifest the role of the bright states in the $\Lambda$ system,
to establish the limits of validity of the simple approximation given
by Eqs. (\ref{eq:Vsubwavelength}) and (\ref{eq:runningapproximation}),
and to find corrections to them. Following the previous discussion,
we consider the control Rabi frequency of the form $\Omega_{c}(z,t)=\Omega_{c}(t)\sin k[z-z_{0}(t)]$,
where $\Omega_{c}(t)$ is periodic with the period $T=2\pi/\omega$.
As a result, the Hamiltonian $H(z,t)$ takes the form
\begin{equation}
H(z,t)=-\frac{\hbar^{2}}{2m_{a}}\partial_{z}^{2}+\hbar\left(\begin{array}{ccc}
0 & \Omega_{c}(z,t)/2 & 0\\
\Omega_{c}(z,t)/2 & -\Delta-i\Gamma/2 & \Omega_{p}/2\\
0 & \Omega_{p}/2 & 0
\end{array}\right)\label{eq:Hamiltonian3levelzt}
\end{equation}
 and is periodic in space and time, $H(z,t)=H(z+\lambda,t)=H(z,t+T)$.
The solutions of the Schrödinger equation 
\begin{equation}
i\hbar\frac{\partial}{\partial t}\psi(z,t)=H(z,t)\psi(z,t)\label{eq:timedependentSchroedinger}
\end{equation}
for the three-component wave function $\psi(z,t)=\{\psi_{g_{1}}(z,t),\,\psi_{e}(z,t),\,\psi_{g_{2}}(z,t)\}$
which defines the state of the system 
\begin{equation}
\ket{\psi(z,t)}=\psi_{g_{1}}(z,t)\ket{g_{1}}+\psi_{e}(z,t)\ket{e}+\psi_{g_{2}}(z,t)\ket{g_{2}},\label{eq:expansion atomic}
\end{equation}
can be taken in the Bloch-Floquet form: 
\begin{equation}
\psi_{q,n,k}(z,t)=e^{iqz}e^{i\epsilon_{q,n,k}t/\hbar}u_{q,n,k}(z,t),\label{eq:quasiperiodicSolutions}
\end{equation}
where $q\in[-\pi/\lambda,\pi/\lambda]$ and $\epsilon_{q,n,k}\in[-\hbar\omega/2,\hbar\omega/2)$
are the quasimomentum and the quasienergy, respectively, and $u_{q,n,k}(z,t)$
is a periodic function both in space and in time, $u_{q,n,k}(z+\lambda,t)=u_{q,n,k}(z,t+T)=u_{q,n,k}(z,t)$.
Here $n$ refers to the Bloch (spatial) band index and $k$ to the
Floquet (temporal) one. Formally, the quasienergies and the corresponding
functions $u_{q,n,k}(z,t)$ can be found by solving the eigenvalue
problem
\begin{equation}
\mathcal{H}_{q}(z,t)u_{q,n,k}(z,t)=\epsilon_{q,n,k}u_{q,n,k}(z,t)\label{eq:eff EV problem}
\end{equation}
 with the Bloch-Floquet Hamiltonian

\begin{eqnarray}
\mathcal{H}_{q}(z,t) & = & H_{q}(z,t)-i\hbar\partial_{t},\label{eq:floquet}
\end{eqnarray}
where 
\begin{equation}
H_{q}(z,t)=e^{-iqz}H(z,t)e^{iqz}=-\frac{\hbar^{2}}{2m_{a}}(\partial_{z}+iq)^{2}+H_{\Lambda}(z,t).\label{eq:Hq}
\end{equation}
After performing the Fourier transform in time, 
\begin{equation}
u_{q,n,k}(z,t)=\sum_{m=-\infty}^{\infty}u_{q,n,k}^{(m)}(z)e^{im\omega t},\label{eq:ueps}
\end{equation}
we can reformulate the eigenvalue problem (\ref{eq:eff EV problem})
as a standard eigenvalue problem
\begin{equation}
\mathcal{\hat{H}}_{q}\mathbf{u}_{q,n,k}=\epsilon_{q,n,k}\mathbf{u}_{q,n,k}\label{eq:eigenvalue problem}
\end{equation}
for the eigenvectors $\mathbf{u}_{q,n,k}=\{u_{q,n,k}^{(m)}(z)\}$
and the Hamiltonian $\mathcal{\hat{H}}_{q}$:

\begin{equation}
\mathcal{\hat{H}}_{q}\mathcal{=\left(\begin{array}{ccccc}
\ddots & H_{1,q} & H_{2,q} & H_{3,q} & H_{4,q}\\
H_{-1,q} & -\omega+H_{0,q} & H_{1,q} & H_{2,q} & H_{3,q}\\
H_{-2,q} & H_{-1,q} & H_{0,q} & H_{1,q} & H_{2,q}\\
H_{-3,q} & H_{-2,q} & H_{-1,q} & \omega+H_{0,q} & H_{1,q}\\
H_{-4,q} & H_{-3,q} & H_{-2,q} & H_{-1,q} & \ddots
\end{array}\right)},\label{eq:cathamiltonian}
\end{equation}
where the operators $H_{n,q}$ originate from the expansion $H_{q}(z,t)=\sum_{s}e^{i\omega st}H_{s,q}(z).$

A version of this approach, which is convenient when the driving frequency
$\omega$ is small compared to the gap $\sim\Omega_{p}$ between the
dark and the bright states and, therefore, when the dark and the bright
channels are only weakly coupled, can be formulated in terms of the
instantaneous dark and bright eigenstates $\{\ket{D(z,t)},\,\ket{B_{+}(z,t)},\,\ket{B_{-}(z,t)}\}$
of the atomic part Hamiltonian (\ref{eq:Hamiltonian3levelzt}), see
Appendix \ref{sec:details} for details. Namely, if we write the state
of the system in the form
\begin{eqnarray}
\ket{\psi(z,t)} & = & \psi^{(d)}(z,t)\ket{D(z,t)}+\psi^{(+)}(z,t)\ket{B_{+}(z,t)}+\nonumber \\
 &  & +\psi^{(-)}(z,t)\ket{B_{-}(z,t)},\label{eq:state expansion instant}
\end{eqnarray}
the Schrödinger equation for the three-component wave function $\tilde{\psi}(z,t)=\{\psi^{(d)}(z,t),\,\psi^{(+)}(z,t),\,\psi^{(-)}(z,t)\}$
reads

\begin{equation}
i\hbar\frac{\partial}{\partial t}\tilde{\psi}(z,t)\!=\!\left[\!-\frac{\hbar^{2}}{2m_{a}}\!\!\left[\partial_{z}+\hat{A}(z,t)\right]^{2}\!\!\!+\hbar\hat{E}(z,t)\!-\!i\hbar\hat{A_{t}}(z,t)\!\right]\tilde{\psi}(z,t),\label{eq:Schr. eq. in instant eigenstates}
\end{equation}
where $\hat{A}$ and $\hat{A}_{t}$ are matrices of the spatial and
temporal connections, respectively, $(\hat{A})_{\alpha\beta}=\bra{\alpha(z,t)}\partial_{z}\ket{\beta(z,t)}$,
$(\hat{A}_{t})_{\alpha\beta}=\bra{\alpha(z,t)}\partial_{t}\ket{\beta(z,t)}$
with $\alpha,\,\beta=D,\,B_{\pm}$, and $\hat{E}(z,t)=\mathrm{diag}[0,\,E_{+}(z,t),\,E_{-}(z,t)]$
(see Appendix \ref{sec:details}). The last term in the above equation
introduce an extra coupling $i\hbar\hat{A_{t}}$ between the dark
and the bright channels, in addition to the coupling $-(\hbar^{2}/m_{a})(\hat{A}\partial_{z}+\partial_{z}\hat{A})$
from the expansion of the kinetic energy term. Another term in this
expansion, $-(\hbar^{2}/2m_{a})\hat{A}^{2}$, gives the nonadiabatic
potential $V_{\mathrm{na}}(z,t)$ in the dark channel. With the ansatz
(\ref{eq:quasiperiodicSolutions}) for the wave function $\tilde{\psi}(z,t)$,
we then end up with the standard eigenvalue problem (\ref{eq:eigenvalue problem}),
(\ref{eq:cathamiltonian}) with the blocks $H_{s,q}(z)$ being now
determined by the Hamiltonian 
\begin{align}
\tilde{H}_{q}(z,t) & =-\frac{\hbar^{2}}{2m_{a}}\left[\partial_{z}+iq+\hat{A}(z,t)\right]^{2}+\hbar\hat{E}(z,t)-i\hbar\hat{A_{t}}(z,t)\label{eq:Htildeq}\\
 & =\sum_{s}e^{i\omega st}H_{s,q}(z).\nonumber 
\end{align}

In the following analysis of the stroboscopic potential painting,
we apply both versions of this general scheme. In Section~\ref{subsec:3finiteDrivingFreq}
we approach the problem from the side of low stroboscopic frequencies
$\omega$ when couplings between the dark and the bright channels
can be ignored. The approach based on the state representation (\ref{eq:state expansion instant})
with the Hamiltonian (\ref{eq:Htildeq}) is then more convenient,
because it allows straightforward reduction of the problem to the
dark channel only and makes numerical simulations simpler. We establish
the lower bound $\omega_{1}$ on $\omega$, above which the approach
based on the time-averaged potential becomes adequate. In Section~\ref{subsec:3DrivingWithBrightState},
we consider higher stroboscopic frequencies $\omega$ and, using the
approach (\ref{eq:expansion atomic}), (\ref{eq:Hq}), analyze the
effects of the bright states on low-energy eigenstates in the dark
channel, as well as establish the upper bound $\omega_{2}$ on $\omega$,
above which the stroboscopic dynamics strongly mixes the dark and
the bright channels. 

\begin{figure}
\includegraphics[width=8.4cm]{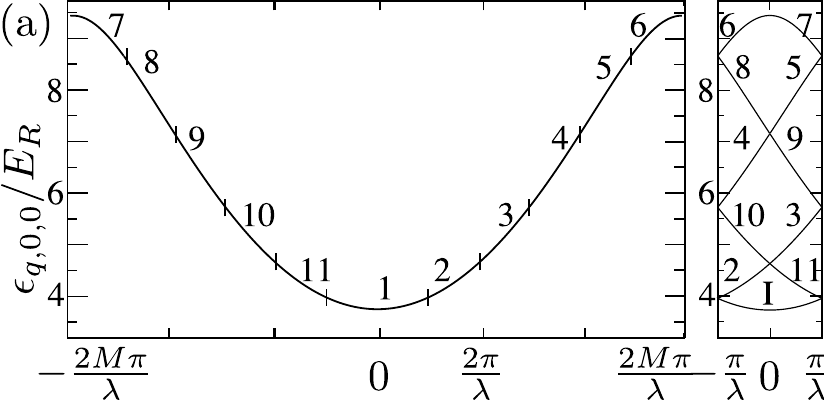}

\includegraphics[width=8.4cm]{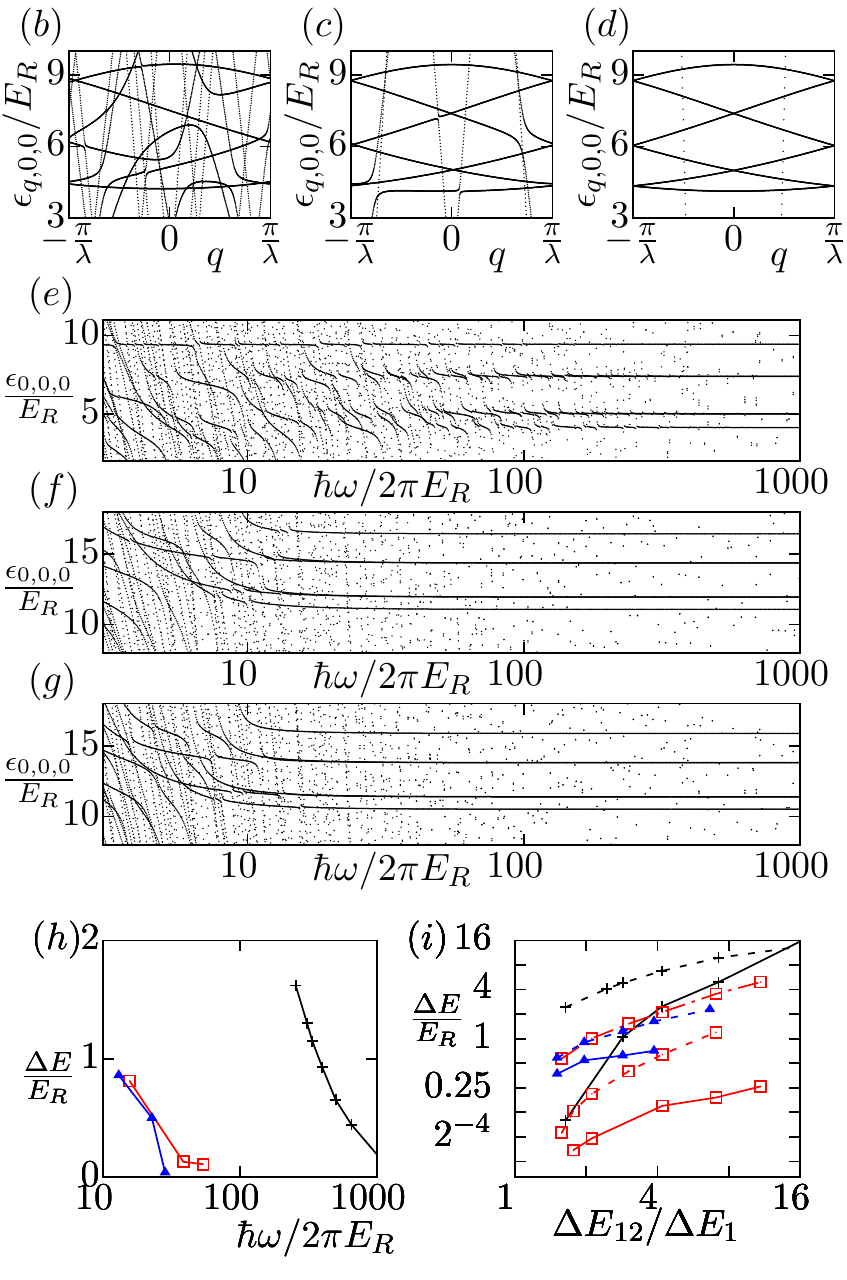}

\caption{Panel (a) shows the lowest band for the potential $\bar{V}(z)$ with
$\epsilon_{0}=1/15$ (see text) in the full Brillouin zone $\mathrm{BZ_{M}}=[-2M\pi/\lambda,2M\pi/\lambda]$
(left) and in the folded Brillouin zone (right). The lowest band for
driving frequencies $\hbar\omega/(2\pi E_{R})=50$ (b), $250$ (c),
and $5000$ (d), with visible avoided crossings for the smallest and
for the intermediate $\omega$. The dependence of the quasienergies
$\epsilon_{q,0,0}$ for $\bar{q}=0,\pi/3\lambda,\ldots,5\pi/3\lambda$
($q=0$ in the folded picture) on the driving frequency $\omega$
for the comb potential painted with Protocol III (e), and for the
smooth potential from Eq.~(\ref{eq:sinpotential}) painted with Protocols
I (f) and II (g) (see text). Panel (h) shows the maximal width of
the avoided crossing $\Delta E$ in the lowest band as a function
of $\omega$: Black crosses are for Protocol III, red squares and
blue triangles for Protocols I and II, respectively. The parameters
are the same as in (e)-(g). Panel (i) compares the dependence of the
size $\Delta E$ of the avoided crossings on the height of the potential,
which is encoded in the ratio of the lowest band gap $\Delta E_{12}$
and the width of the lowest band $\Delta E_{1}$, for different protocols
(see text). Triangles refer to Protocol I (solid line for $\omega=2\pi\times22E_{R}/\hbar$
and dashed for $\omega=2\pi\times13E_{R}/\hbar$ ), rectangles to
Protocol II (solid for $\omega=2\pi\times62.5E_{R}/\hbar$, dashed
for $\omega=2\pi\times22.7E_{R}/\hbar$, and dash-dotted for $\omega=2\pi\times16.1E_{R}/\hbar$),
and crosses to the Protocol III (solid for $\omega=2\pi\times910E_{R}/\hbar$
and dashed for $\omega=2\pi\times161E_{R}/\hbar$).}

\label{fig:bandStructure}
\end{figure}

\section{Dark state regime }

\label{subsec:3finiteDrivingFreq} 

We first consider the case when the stroboscopic frequency $\omega$
is much smaller that the energy gap to the bright states in the atomic
$\Lambda$ system, such that the dynamics occurs entirely within the
dark state and the couplings to the bright states can be ignored.
We can therefore keep only the dark state part in the decomposition
(\ref{eq:state expansion instant}) and in the equation (\ref{eq:Schr. eq. in instant eigenstates}).
Within the ansatz (\ref{eq:quasiperiodicSolutions}) for the dark
state wave function, the term $H_{q}$ in the Bloch-Floquet Hamiltonian,
Eq.~(\ref{eq:floquet}), then reads

\begin{equation}
H_{q}(z,t)=-\frac{\hbar^{2}}{2m_{a}}\left(\partial_{z}+iq\right)^{2}+V_{\textrm{na}}[\epsilon(t),z-z_{0}(t)],\label{eq:hamiltonianq}
\end{equation}
where $\epsilon(t)$ and $z_{0}(t)$ correspond to the chosen stroboscopic
protocol. 

A time averaging of $H_{q}(z,t)$ over the stroboscopic period gives
the operator $H_{0,q}(z)$ that appears on the diagonal in Eq.~(\ref{eq:cathamiltonian}).
We note that the operator $H_{0,q}$ contains the time-averaged potential
$\bar{V}(z)$ which is periodic and, therefore, the equation $H_{0,q}(z)u_{q,n}(z)=\epsilon_{q,n}^{(0)}u_{q,n}(z)$
determines the eigenenergies of the Bloch states in the potential
$\bar{V}(z)$ for a given quasimomentum $q$, where $n$ is the Bloch
band index. For a potential $\bar{V}(z)$ which has $M$ times shorter
period $a=\lambda/(2M)$ than $V_{\textrm{na.}}(z)$, the Brillouin
zone $q\in[-\pi/a,\,\pi/a]=[-kM,\,kM]\equiv\mathrm{BZ_{M}}$ is $2M$
times larger than the Brillouin zone $q\in[-\pi/\lambda,\,\pi/\lambda]=[-k/2,\,k/2]\equiv\mathrm{BZ}$
for solutions of Eq.~(\ref{eq:eff EV problem}), which are $\lambda$
periodic. The two Brillouin zones can be mapped onto each other by
introducing a sub-band index $n'=0,1,\ldots,\,2M$ according to the
rule $\overline{q}=q+\pi n'\mathrm{sign}(q)/\lambda=q+kn'\mathrm{sign}(q)/2$,
where $\overline{q}\in\mathrm{BZ_{M}}$ and $q\in\mathrm{BZ}$. Fig.~\ref{fig:bandStructure}($a$)
shows an example of such mapping for the quasienergy $\epsilon_{q,0,0}$
in the lowest ($n=0$ and $k=0$) Bloch-Floquet band obtained by using
Protocol III for the comb potential with $M=3$. 

When one neglects the off-diagonal blocks in the Hamiltonian (\ref{eq:cathamiltonian}),
the quasienergies will simply be $\epsilon_{q,n,m}^{(0)}=\epsilon_{q,n}^{(0)}+\hbar\omega m$.
Non-diagonal blocks $H_{m,q},\,m\neq0$ couple the bare states in
different Floquet and Bloch bands via the matrix elements $(H_{m,q})_{n_{1}n_{2}}$.
This coupling is essential when the two coupled level are close in
energy, $\epsilon_{q,n_{1},m_{1}}^{(0)}\thickapprox\epsilon_{q,n_{2},m_{2}}^{(0)}+\hbar\omega m$,
and results in the avoiding crossing with the gap $\Delta E=2\left|(H_{m,q_{0}})_{n_{1}n_{2}}\right|$,
where $q_{0}$ corresponds to the resonance, $\epsilon_{q_{0},n_{1},m_{1}}^{(0)}=\epsilon_{q_{0},n_{2},m_{2}}^{(0)}+\hbar\omega m$.
For low-energy bands $\epsilon_{q,n,0}^{(0)}=\epsilon_{q,n}^{(0)}$
of $H_{0,q}$ (with small $n$), a level $\epsilon_{n_{1},q}^{(0)}$
with $n_{1}\sim1$ can be significantly coupled to the level $\epsilon_{q,n_{2},m}^{(0)}=\epsilon_{n_{2},q}^{(0)}+m\omega\hbar$
provided $\left|\epsilon_{n_{1},q}^{(0)}-\epsilon_{n_{2},q}^{(0)}-\hbar\omega m\right|\lesssim\left|(H_{m,q})_{n_{1}n_{2}}\right|$.
In the considered case of a large driving frequency $\hbar\omega\gg E_{R}$,
this takes place for $m<0$ and $n_{2}\sim\sqrt{m\hbar\omega/E_{R}}\gg1$,
meaning the coupling to the states in the high bands where $\epsilon_{q,n}^{(0)}$
generically changes rapidly with $q$. As a result, the range $\delta q$
of the quasimomentum where the coupling is important can be estimated
as $\delta q\sim(H_{m,q})_{n_{1}n_{2}}/v_{n_{2}}(q)$, where $v_{n_{2}}(q)=d\epsilon_{n_{2},q}^{(0)}/dq$
is the group velocity in the band $\epsilon_{n_{2},q}^{(0)}$ at the
resonant quasimomentum. Keeping in mind that the matrix elements $(H_{m,q})_{n_{1}n_{2}}$
decay fast with $|n_{2}-n_{1}|\sim\sqrt{\omega}$, we conclude that
the low-energy part of the coupled band structure is given by the
bands of $\bar{V}$ with avoided crossings which widths decay with
$\omega$, see Fig.~\ref{fig:bandStructure}. 

To support this picture, we performed numerical simulations of the
the Floquet problem for all three protocols considered above. For
this purpose we truncate the matrix in Eq.~(\ref{eq:cathamiltonian})
up to $|m|\leq10$ and $\left|n_{z}\right|<300$ spatial Fourier components
in the decomposition of the eigenvectors $\mathbf{u}_{q,n,k}(z)=\sum_{n_{z}}\mathbf{u}_{q,n,k}(n_{z})\exp(ikn_{z}z)$
(we have checked that increasing these cutoffs to larger values does
not change the results), and use the SciPy Python library \cite{Bressert2012}
to compute the eigenvalues. Figures.~\ref{fig:bandStructure}(b)-(d)
show the results for the quasienergy $\epsilon_{q,0,0}$ in the first
Bloch-Floquet band (in the folded Brillouin zone picture) for the
comb potential with $\epsilon_{0}=1/15$ and $M=3$, calculated using
Protocol III (see details in Appendix~\ref{sec:AppTimeDependence})
for driving frequencies $\hbar\omega/E_{R}=2\pi\times50$, $2\pi\times250$,
and $2\pi\times5000$, respectively. For the smallest driving frequency,
the couplings to high bands disrupt the band structure almost completely.
With increasing the driving frequency to the intermediate value, the
effect of the couplings decreases drastically but is still visible,
and for the largest $\omega$ the lowest band looks practically perfect,
with only a few extremely narrow avoided crossing regions (see however
the discussion of the effects of the couplings to the bright states
in Section~\ref{subsec:3DrivingWithBrightState}). The calculated
parameters of the low-energy band structure in this case are: $\textrm{max }\bar{V}(z)=63E_{R}$,
bandwidth of the lowest band of $\Delta E_{1}\approx5.3E_{R}$, and
the band gap to the next band $\Delta E_{12}\approx8.7E_{R}$, such
that $\Delta E_{12}/\Delta E_{1}\approx1.64$.

The band structures calculated for the potential in Eq.~(\ref{eq:sinpotential})
with $M=3$ painted by using the Protocols I and II, show the same
features with only some minor quantitative differences. To compare
different protocols, we chose the value of the amplitude $V$ in Eq.~(\ref{eq:sinpotential})
to be $V=17.2E_{R}$ for the Protocol I and $V=18.2E_{R}$ for the
Protocol II ($W=3E_{R}$ in both cases). With these values, for both
protocols we obtain $\Delta E_{1}=5.3E_{R}$ for the width of the
lowest band and $\Delta E_{12}=8.8E_{R}$ for the gap between the
lowest bands, such that the ratio $\Delta E_{12}/\Delta E_{1}\approx1.66$
is almost the same as in the above case of the comb potential with
$M=3$ painted with the Protocol III. 

The results of the three protocols are compared in Figs.~\ref{fig:bandStructure}(e)-(g)
where we present the dependence of the lowest band quasienergy $\epsilon_{q,0,0}$
for the quasimomenta $\bar{q}=0,\pi/3\lambda,\ldots,5\pi/3\lambda$
(corresponding to the folded quasimomentum $q=0$) on the driving
frequency $\omega$ for the Protocols III, I, and II, respectively.
For each $\omega$ we have 6 values of $\epsilon_{q,0,0}$ immersed
in the background of the quasienergies from other Floquet bands, which
show a complicated avoided crossing structure for small $\omega$
but form a continuous lines (two middle ones are double degenerate)
when $\omega$ increases and the description in terms of the averaged
potential $\bar{V}(z)$ becomes accurate. We see that in the case
of the comb potential {[}Fig.~\ref{fig:bandStructure}(e){]}, the
region of significant avoided crossings extends to much high driving
frequencies as compared to the cases of the smooth potential {[}Figs.~\ref{fig:bandStructure}(f)
and (g){]}. This is a result of a much slower decay of the Fourier
components, both in frequency and in momentum spaces, of the $V_{\mathrm{na}}(z,t)$
in this case.

To compare the avoided crossings for different protocols quantitatively,
we consider those values of $\omega$ when the avoided crossing takes
place for $\bar{q}=0$ (at the bottom of the Bloch band). The corresponding
avoided crossing gaps $\Delta E$ for these values of $\omega$ are
plotted in Fig.~\ref{fig:bandStructure}(h) for all three protocols,
showing that the gaps $\Delta E$ for the smooth Protocols I and II
are generically much smaller than that for the Protocol II, and vanishes
much faster with increasing $\omega$. In Figure~\ref{fig:bandStructure}(i)
we present for several values of $\omega$ and different painting
protocols the dependence of the maximal $\Delta E$ across the lowest
band on the ratio $\Delta E_{12}/\Delta E_{1}$ which is mostly determined
by the height of the painted potential. To obtain these dependence,
we scan the parameters used for painting ($\epsilon_{0}$ for Protocol
III and both $V$ and $W$ for Protocols I and II) such that the painted
potentials have band structures with the ratio $\Delta E_{12}/\Delta E_{1}$
in the range $1-20$. The values of the stroboscopic frequency $\omega$
are chosen in the range where the values of $\Delta E$ are between
a fraction of $E_{R}$ and a few $E_{R}$. For the smooth Protocols
I and II this results in the values of $\omega$ between $2\pi\times13E_{R}/\hbar\textrm{ and }2\pi\times67.5E_{R}/\hbar$,
and between $2\pi\times161E_{R}/\hbar$ and $2\pi\times552E_{R}/\hbar$
for Potential III. Again we see that painting of a comb potential
is generically much more demanding to the value of the stroboscopic
frequency $\omega$, in order to produce the lowest band of the same
``quality''. Note that the growth of $\Delta E$ with $\Delta E_{12}/\Delta E_{1}$
for a fixed $\omega$, which is clearly seen in Fig.~\ref{fig:bandStructure}(i),
is mostly related to the increase of the coupling matrix element $(H_{1,q})_{0n_{2}}$
which couples the lowest Bloch band with the high ones.

\section{Bright states effects}

\label{subsec:3DrivingWithBrightState}

\begin{figure}
\includegraphics[width=8.6cm]{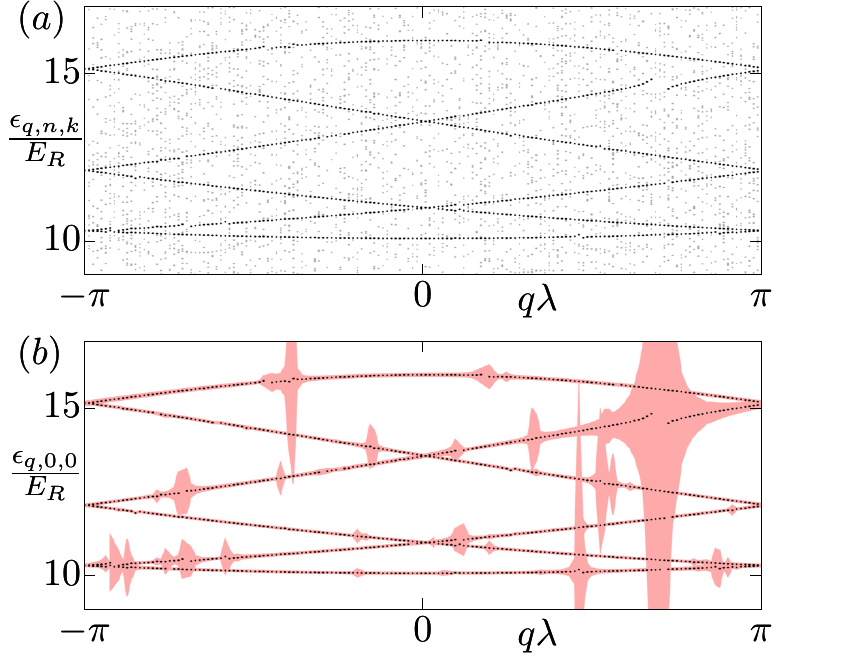}

\caption{Low energy Bloch-Floquet eigenenergies calculated using Eqs.~(\ref{eq:cathamiltonian})
and (\ref{eq:Hamiltonian3levelzt}) for the smooth potential, Eq.~(\ref{eq:sinpotential})
with $W=3E_{R}$ and $V=15E_{R}$, painted using Protocol II. Panel
(a) shows real part $\textrm{Re}\epsilon_{q,0,0}$ of the quasienergies,
with black dots used for the states in the lowest Bloch-Floquet band
and gray dots for all others. In panel $(b)$ we indicate also the
imaginary part $\textrm{Im}\epsilon_{q,0,0}$ of the quasienergies
in the lowest band by the shaded region with boundaries $\textrm{Re}\epsilon_{q,0,0}\pm\textrm{Im}\epsilon_{q,0,0}$.
The data in both panels are for $\omega=2\pi\times30E_{R}/\hbar$
and $\Omega_{c}=2\times10^{6}E_{R}$ ($\epsilon_{0}=0.06$). The width
of the excited atomic state $\ket{e}$ is $\Gamma=300E_{R}.$}

\label{fig:complexQuasienergies}
\end{figure}

\begin{figure}
\includegraphics[width=8.6cm]{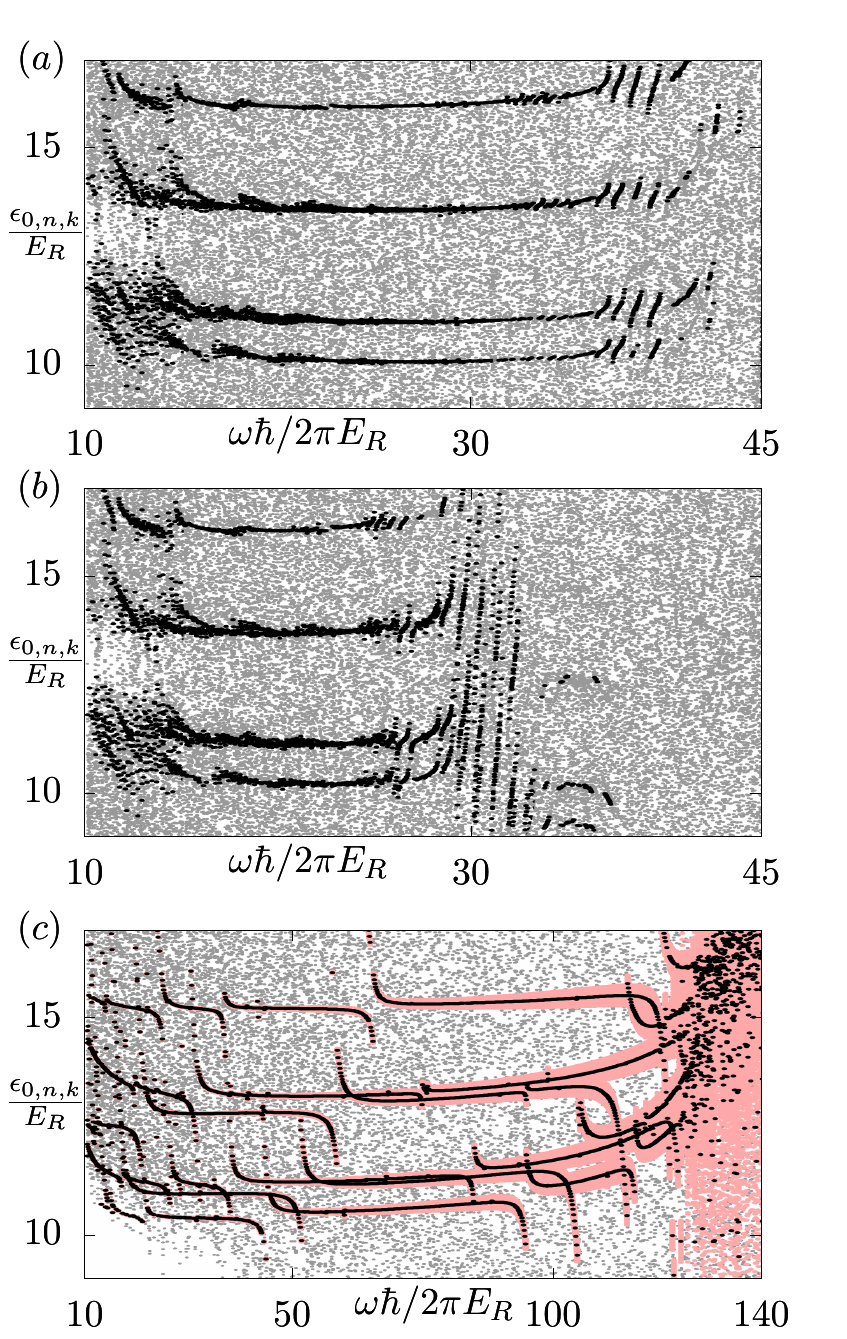}

\caption{Low-energy Bloch-Floquet eigenenergies for quasimomenta $\bar{q}=0,\pi/3\lambda,\ldots,5\pi/3\lambda$
($q=0$ in the folded picture) as a function of the driving frequency
$\omega$, calculated on the basis of Eqs.~(\ref{eq:cathamiltonian})
and~(\ref{eq:}) (see text). Panels (a) and (b) show real parts of
the quasienergies when Protocol II is used with $\Omega_{c}=2\times10^{6}E_{R}/\hbar$
and $\Omega_{c}=1.5\times10^{6}E_{R}/\hbar,$respectively, ($\epsilon_{0}=0.06$
in both cases) for painting the smooth potential given by Eq. (\ref{eq:sinpotential})
with $W=3E_{R},\,V=15E_{R}$, and $M=3$. The Panel $(c)$ is for
the comb potential with $M=3$ painted with Protocol III when $\epsilon_{0}=0.2$
and $\mathrm{max_{z}}\Omega_{c}=2\times10^{6}E_{R}/\hbar$, where
we also indicate the imaginary parts for the states in the lowest
Bloch-Floquet band by the shaded region with boundaries $\textrm{Re}\epsilon_{q,0,0}\pm\textrm{Im}\epsilon_{q,0,0}$,
calculated for $\Gamma=300E_{R}.$}

\label{fig:omegadependence3c}
\end{figure}
From the results of the previous Section we see that the stroboscopic
approach in the single-channel (dark-state) approximation becomes
increasingly accurate with increasing the driving frequency $\omega$.
However, one would expect that for $\omega$ of the order of the gap
between the dark and the bright states (actually, much earlier due
to higher Fourier components, see below), couplings to the bright
states become important and, therefore, the single-channel approach
of the previous Section fails. Another effect of the bright states
which shows up before the driving frequency becomes comparable to
the gap, is the appearance of significant imaginary parts for the
quasienergies in the lowest dark-state band in the regions of avoided
crossing. These imaginary parts are much larger than one would expect
on the basis of the direct coupling between the states in the lowest
dark-state band and the bright states \cite{Lacki2016}. They originate
from the large imaginary parts for the quasienergies of the states
in the high-energy dark-state band (having much higher energy, they
are stronger coupled to the bright states) which strongly coupled
to the states in the lowest dark-state band in the avoided crossing
regions.

To take the bright states into account, we consider the Floquet problem
with the Hamiltonian (\ref{eq:cathamiltonian}), where the Floquet
blocks $H_{s,q}(z)$ are determined by the full Hamiltonian (\ref{eq:Hamiltonian3levelzt}).
For our numerical calculations we consider the cases of a smooth and
of a comb potentials, both with the spatial period $\lambda/6$ ($M=3$).
The smooth potential is taken in the form (\ref{eq:sinpotential})
with $W=3E_{R}$ and $V=15E_{R}$, and for its painting we use Protocol
II with $\epsilon_{0}=0.06$. In painting the comb potential according
to Protocol III, we take $\epsilon_{0}=0.2$ (see below) and fix the
maximal value of $\Omega_{c}(z,t)$ to $2\times10^{6}E_{R}/\hbar$
(see also Appendix \ref{sec:AppTimeDependence}) which corresponds
to $\Omega_{p}=4\times10^{5}E_{R}/\hbar$.

The Floquet-Bloch Hamiltonian $\mathcal{\hat{H}}_{q}$, Eq.~(\ref{eq:cathamiltonian}),
is truncated to $|m|\leq300$ and $|n_{k}|<300$ (we get similar results
with twice larger cutoffs), and in calculating its matrix elements
we use 2D Discrete Fourier Transform and keep only terms which are
larger than $10^{-6}\max_{i,j}|(\mathcal{H}_{q})_{i,j}|.$ This gives
$\sim100$ nonzero entries per row for Protocol II and $604$ for
Protocol III in the $1\text{ }083\text{ }603\times1\text{ }083\text{ }603$
matrix. By using Sparse diagonalization routines from Python's SciPy
library \cite{Bressert2012}, we extract 30 eigenvalues which are
the closest to the lowest dark-state band in the $m=0$ Floquet block.

Figure~\ref{fig:complexQuasienergies}(a) shows the real parts of
the quasienergies in the lowest band for the case of the smooth potential
painted with the driving frequency $\omega=2\pi\times30E_{R}/\hbar$
which is well below the gap $\Omega_{p}/2=6.7\times10^{4}E_{R}/\hbar$
to the bright state, such that direct couplings between dark states
in the lowest band and bright states are very small. We see a well-defined
band with few avoided crossings due to couplings to the states in
the high dark-state bands. This band is immersed into the background
of other quasienergies which we plot only if the corresponding eigenstates
have overlap more than $0.7$ with the eigenstates $\psi_{q,n,0}^{(d)}$
calculated for the dark channel only (see previous Section) in the
limit $\omega\to\infty$ (with the time averaged potential). In Fig.~\ref{fig:complexQuasienergies}(b)
we show the corresponding imaginary part which are generically very
small due to small couplings to the bright states (see \cite{Lacki2016,Wang2017}),
except for several regions with avoided crossings. As it was pointed
out before, these large imaginary parts are due to large imaginary
parts of the admixed high-energy dark states from different Floquet
blocks, which are strongly coupled to decaying bright states.

In Fig.~\ref{fig:omegadependence3c} we present the quasienergies
for $q=0$ as a function of driving frequency $\omega$ for the smooth
case with Protocol II {[}panels $(a)$ and $(b)${]} and for the comb
case with Protocol III {[}panel $(c)${]}. The black points mark quasienergies
in the lowest Floquet block, which correspond to the eigenstates with
an overlap of at least $0.99$ with the dark channel eigenstates $\psi_{q,n,0}^{(d)}$.
Other quasienergies are gray points and form a background. In the
region of small $\omega,$ we recover the same behavior as in Figs.
4(e)-(g) with formation of a well-defined dark-state Bloch band for
the driving frequencies $\omega$ larger than some typical value $\omega_{1}$.
For the entire system, however, with increasing $\omega$, the dark-state
band structure terminates at some value $\omega_{2}$ due to strong
coupling to the bright states, when also substantial imaginary parts
appear {[}see Fig. 6(c), similar behavior is also present for the
cases in panels (a) and (b) but is not shown{]}. The value of $\omega_{2}$
is related to the gap $\Omega_{p}/2$ as it is demonstrated in Figs.
6 (a) and (b) which corresponds to the same $\epsilon_{0}=0.06$ but
different $\Omega_{c}$ and, therefore, different $\Omega_{p}$: $\Omega_{p}=1.2\times10^{5}E_{R}/\hbar$
for $\Omega_{c}=2\times10^{6}E_{R}/\hbar$ and $\Omega_{p}=9\times10^{4}E_{R}/\hbar$
for $\Omega_{c}=1.5\times10^{6}E_{R}/\hbar$, respectively. Note,
however, that $\omega_{2}$ is typically $2$ orders of magnitude
smaller that the gap $\Omega_{p}/2$, indicating the importance of
higher Fourier components $m\omega$ in the couplings to the bright
states. We mention also that in the case of Protocol III, Fig.~\ref{fig:complexQuasienergies}(c),
the chosen value $\epsilon_{0}=0.2$ results in a rather shallow comb
potential with the height $\sim7E_{R}.$ This, together with rapid
change of $\Omega_{c}(z,t)$ in time and space and, hence, the presence
of high Fourier components, results in a stronger coupling to the
bright states for this Protocol. Note that, if for Protocol III we
decrease $\epsilon_{0}$ without increasing $\mathrm{max_{z}}\Omega_{c}(z)$,
the result will be the decrease of the gap $\Omega_{p}/2$ to the
bright states, and, as a consequence, the destruction of the low-energy
bands in the dark channel. Taking smaller values of $\epsilon_{0}$
would therefore require considering much high values of $\Omega_{c}$
which is problematic for numerical simulations.

The above examples show that the stroboscopic painting of a smooth
potential requires less restrictive conditions on the Rabi frequencies
$\Omega_{c}$ and $\Omega_{p}$ to ensure the existence of an intermediate
driving frequency regime where the description based on the dark-state
channel with a with time-averaged potential is legitimate. This is,
of course, related to the fact that the smoother the function, the
faster high Fourier components decay. Therefore, for a rapidly varying
potential one needs a much larger gap to the bright states, as compared
to a smooth potential, in order to minimize their effect on the low-energy
states in the dark channel.

\section{Bloch oscillations}

\begin{figure}
\includegraphics[width=8.4cm]{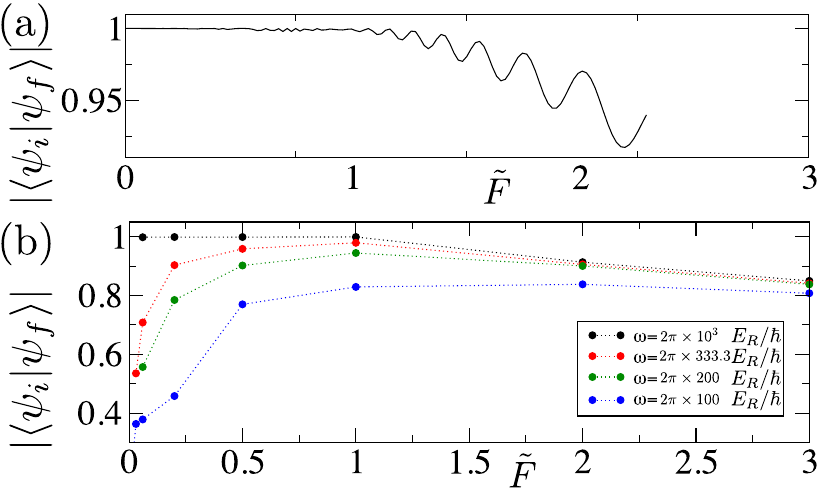}

\caption{The overlap of initial Bloch state $\psi_{i}$ on the final state,
after quasimomentum corresponding to the BZ width has been slowly
imprinted . Panel $(a)$ shows such fidelity $\mathcal{F}=|\langle\psi_{i}|\psi_{f}\rangle|$
in case of evolution in tilted time-averaged potential $\bar{V}$,
i.e no avoided crossings present, $\omega\to\infty$ limit in the
dark state channel. Drop of the fidelity for $\tilde{F}>\tilde{F}_{0}$
due to population of higher Bloch bands. Panel (b) shows $\mathcal{F}$
computed for the dark state Floquet Hamiltonian, for the comb potential
$\epsilon_{0}=2/30,N=3$ with $M=5$ Floquet blocks for several values
of projection frequency$\omega$. For $\omega\to\infty$ limit from
$(a)$ is recovered. For finite $\omega$ the fidelity is maximized
for the moderate tilt of the lattice $\tilde{F}.$ For $\tilde{F}\to0$
the state follows adiabatically any avoided crossing, for $\tilde{F}\to\infty$
population is spread to higher Bloch bands.-}

\label{fig:5BlochOscillationsFidelity}
\end{figure}
\label{sec:BlochOscillations}As we have seen in the previous Section,
the low-energy band structure of the painted potentials has a number
of avoided crossings due to coupling to states in high-energy bands.
For an evolving state in the lowest band, these avoided crossings
result in a ``leakage'' into higher-energy states each time the
evolving state passes through them. Intuitively, the faster the speed
at which the state passes through the avoided crossing, the smaller
the fraction transferred into the higher bands. On the other hand,
applying too strong forces in order to speed up the dynamics can also
introduce transitions into higher bands. We address these effects
below by considering Bloch oscillations in the lowest band. Our consideration
will be limited to the dark-state channel, and the effects of the
bright states will be discussed at the end of the Section.

To initiate the Bloch oscillations, we apply a constant force $F$
which corresponds to a linear potential $-Fz$ added to the Hamiltonian.
This force results in an evolution of the quasimomentum $q(t)=q_{i}+Ft/\hbar$
which, starting from an initial value $q_{i}$, winds around the Brillouin
zone after the Bloch oscillation period $\tau_{B}$, $q(\tau_{B})=q_{i}+2\pi/\lambda$
with $\tau_{B}=2\pi\hbar/(F\lambda)=\hbar k/F$. As an initial state
$\ket{\psi_{i}}$ we take a state with a quasimomentum $q_{i}$ in
the lowest Bloch-Floquet band, and apply the the time-evolution operator
$\mathcal{T}\exp\left[i/\hbar\int(H_{q(t)}-Fz)\textrm{d}t\right]$
during time $\tau_{B}$ with $H_{q}$ from Eq.~(\ref{eq:hamiltonianq}).
The resulting final states $\ket{\psi_{f}}$ of the system is then
compared with the initial one by calculating the fidelity $\mathcal{F}=|\langle\psi_{i}|\psi_{f}\rangle|$.
The decrease of $\mathcal{F}$ from its maximal value $1$ determines
the loss into higher bands. For our numerical simulations we use Protocol
III to paint a comb potential with $M=3$ peaks per length $a=\lambda/2$
and $\epsilon_{0}=1/15$. We choose this protocol because it results
in the most prominent avoided-crossing structure in the lowest band
among the three considered protocols. The calculation of the time
evolution operator is performed by Lanczos algorithm implemented with
adaptive time-steps. 

Let us first establish the upper bound on the force $F$. For this
purpose we consider the limit $\omega\to\infty$ when the Hamiltonian
$H_{q}$ contains the time-averaged potential $\bar{V}(z)$ and no
avoided crossings are present in the band structure. Figure~\ref{fig:5BlochOscillationsFidelity}(a)
shows the fidelity $\mathcal{F}$ for this case as a function of the
dimensionless force $\tilde{F}=F/(kE_{R})=\hbar/(\tau_{B}E_{R}).$
We see that the fidelity $\mathcal{F}$ shows noticeable decrease
for $\tilde{F}\geq\tilde{F}_{0}\sim1$ due to inter band transitions,
when the change of the potential energy caused by the force $F$ over
the lattice period, $\Delta E\sim F\lambda$, becomes of the order
of the gap between the lowest and the next Bloch bands. We therefore
obtain the condition $F<kE_{R}$ which sets the upper bound on the
force $F$ and ensures the absence of the interband transitions in
the case of no avoided crossings. 

For a finite stroboscopic frequency $\omega$, we perform our calculations
using the Floquet scheme with the Hamiltonian~(\ref{eq:hamiltonianq}).
In these calculations, for each set of parameters we chose an initial
quasimomentum $q_{i}$ to be far away from any avoided crossing. The
calculated dependence of the fidelity ${\cal F}$ on the dimensionless
force $\tilde{F}$ is shown in Fig.~\ref{fig:5BlochOscillationsFidelity}(b)
for different $\omega$. As in Fig.~\ref{fig:5BlochOscillationsFidelity}(a),
we see a decrease of ${\cal F}$ for $\tilde{F}>1$ for all $\omega$
due to interband transitions. The behavior of the fidelity for $\tilde{F}<1$
strongly depends on the the value of $\omega$. For the highest considered
frequency $\omega=2\pi\times10^{3}E_{R}/\hbar$, the fidelity $\mathcal{F}$
remains very close to unity even for very small values of the force
$\tilde{F}$, while we see a strong decrease of the fidelity with
decreasing $\tilde{F}$ for smaller values of $\omega$. This behavior
is related to the $\omega$-dependence of the widths of the avoided
crossing regions, which grow with decreasing $\omega$ {[}see discussion
in Sec.~\ref{subsec:3finiteDrivingFreq} and Fig.~\ref{fig:bandStructure}(h){]}.
As a result, the higher the stroboscopic frequency $\omega,$ the
more diabatic is the traverse through the avoided crossing region
for a given force $\tilde{F}$.

A more quantitative picture can be obtained on the basis of the Landau-Zener
expression for the probability of non-adiabatic passage during a linear
sweep through a single avoided crossing \cite{Kleppner1981} 
\begin{equation}
P=\exp(-2\pi\Gamma),\quad\Gamma=\frac{|\Delta E/2|^{2}}{\alpha\hbar},\label{eq:avcros}
\end{equation}
where $\Delta E$ is the minimal energy difference (twice the coupling
matrix element) in the avoided crossing and $\alpha$ is the speed
with which the energy difference between the crossing levels changes
during the sweep. To estimate $\alpha$ we assume that the avoided
crossing takes place at some quasimomentum $q_{*}$, when two states,
one in the lowest band ($n=0$) and one in the high-energy band ($n_{1}\gg1$)
are at resonance, $\epsilon_{q_{*},n_{1},0}^{(0)}-\epsilon_{q_{*},0,0}^{(0)}=\hbar\omega$,
where we consider the Floquet resonance with $m=1$ because it results
in the strongest coupling matrix element. For the dispersion $\epsilon_{q,n_{1},0}^{(0)}$
in the high-energy band we can assume a quadratic in form $\epsilon_{q,n_{1},0}^{(0)}\approx\hbar^{2}(q+2kMn_{1})^{2}/2m$
and, therefore, for the group velocity at $q=q_{*}$ we obtain $v_{n_{1}}(q_{*})=d\epsilon_{q_{*},n_{1},0}/dq\approx\hbar\sqrt{2\hbar\omega/m}$.
As a result, the energy difference between the crossing levels during
the sweep can be written as $\Delta\epsilon(q)=\epsilon_{q,n_{1},0}^{(0)}-\epsilon_{q,0,0}^{(0)}-\hbar\omega\approx\epsilon_{q,n_{1},0}^{(0)}-\hbar\omega\approx v_{n_{1}}(q_{*})(q-q_{*})$
for $q\approx q_{*}$, and for the speed $\alpha$ we obtain $\alpha=v_{n_{1}}(q_{*})F/\hbar=F\sqrt{2\hbar\omega/m}$.
The expression for $\Gamma$ can now be written in the form
\[
\Gamma=\frac{|\Delta E|^{2}}{\hbar F\sqrt{2\hbar\omega/m}}=\left|\frac{(H_{1,q_{*}})_{0,n_{1}}}{E_{R}}\right|^{2}\frac{1}{2\tilde{F}\sqrt{\hbar\omega/E_{R}}},
\]
and we see that a substantial adiabatic passage through the avoided
crossing and, therefore, strong loss into high bands, occurs when
$\tilde{F}\lesssim\tilde{f}_{0}\sim[(H_{1,q_{*}})_{0,n_{1}}/E_{R}]{}^{2}\sqrt{E_{R}/\hbar\omega}$.
The value of $\tilde{f}_{0}$ decreases rapidly with increasing $\omega$,
also due to the decrease of the coupling matrix element $(H_{1,q_{*}})_{0,n_{1}}$
for large $n_{1}\sim\sqrt{\omega}$, in accordance with Fig.~\ref{fig:5BlochOscillationsFidelity}(b).

Let us now briefly discuss the effects of the bright channels. It
is clear, that the above consideration of the Bloch oscillations makes
sense only when the couplings to the bright channels are small such
that the dark-channel low-energy bands are well defined, that is for
$\omega$ much less that the gap to the bright states. In this case,
the effect of the bright states can be taken into account by introducing
an effective overall decay rate $\Gamma_{\mathrm{eff}}(\omega)$ which
is an averaged (over the Brillouin zone) decay rate of the states
in the lowest-energy Bloch band in the dark channel, see Fig.~\ref{fig:complexQuasienergies}(b).
The fidelity $\mathcal{F}$ gets then an extra factor $\exp[-\Gamma_{\mathrm{eff}}(\omega)\tau_{B}]$
reflecting a finite life-time of the states in the dark channel. Note
that the quantity $\Gamma_{\mathrm{eff}}(\omega)$ can be computed
within the three-channel calculations from Sec.~\ref{subsec:3DrivingWithBrightState},
similar to the approach used in \cite{Wang2017} for describing an
effective loss rate in the experiments with a static comb potential
implemented using an atomic $\Lambda$ scheme. 

\section{Loading protocol}

\begin{figure}
\includegraphics[width=8cm]{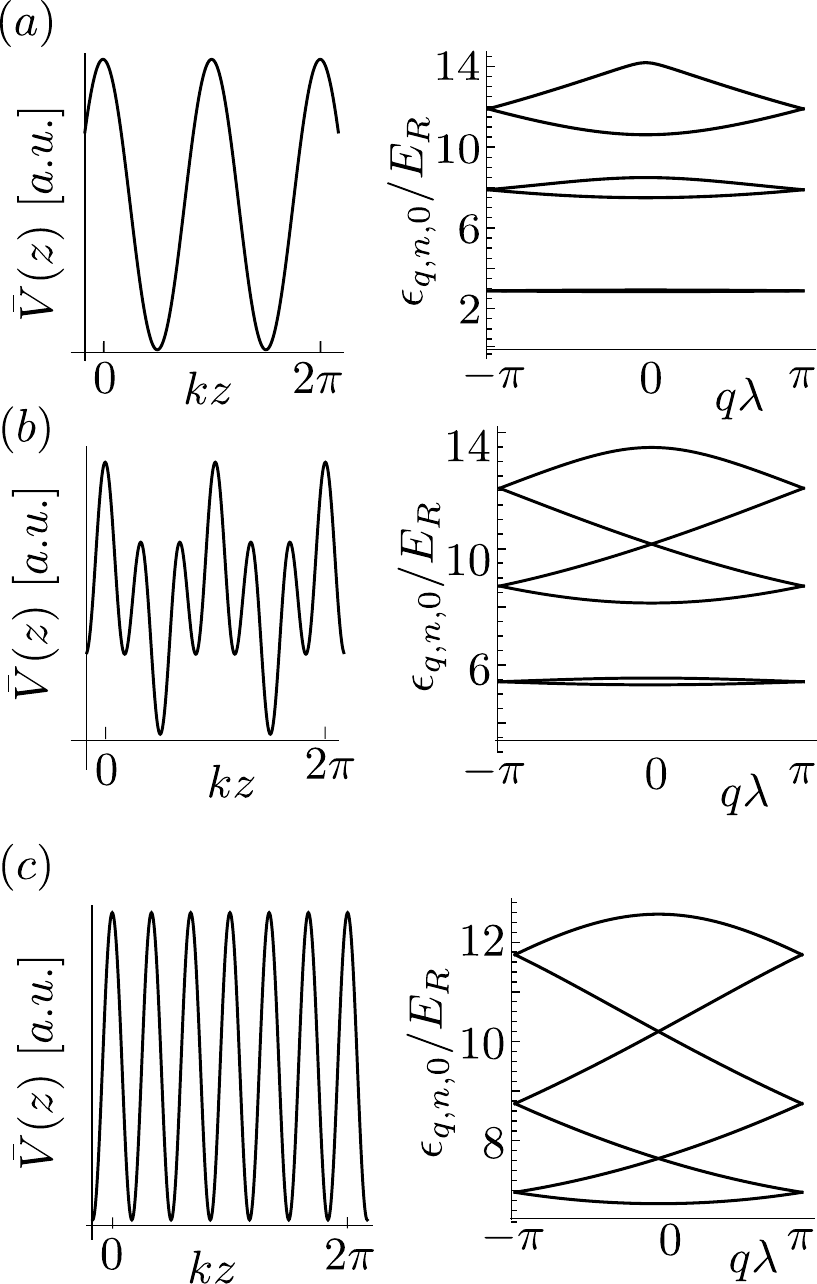}

\caption{Steps of the protocol for loading an atom into the lowest band of
the time-averaged potential $\bar{V}(z)$ from Eq.~(\ref{eq:sinpotential}).
Each panel shows the time-averaged potential on the left part and
the corresponding band structure on the right part. Panel $(a)$ shows
the beginning of the loading, when the atom is in the lowest band
of the auxiliary time-independent optical lattice potential $V_{L}(z)$,
and also the end of the first step if all plots are shifted up in
energy by $W$. During the second step we slowly switch $V_{L}(z)$
off together with switching on the amplitude $V$ in the painted potential.
An intermediate moment of the second step is presented in panel (b),
and the final moment in panel (c) (see text).}

\label{fig:loadingsmooth}
\end{figure}

\begin{figure}
\includegraphics[width=8.4cm]{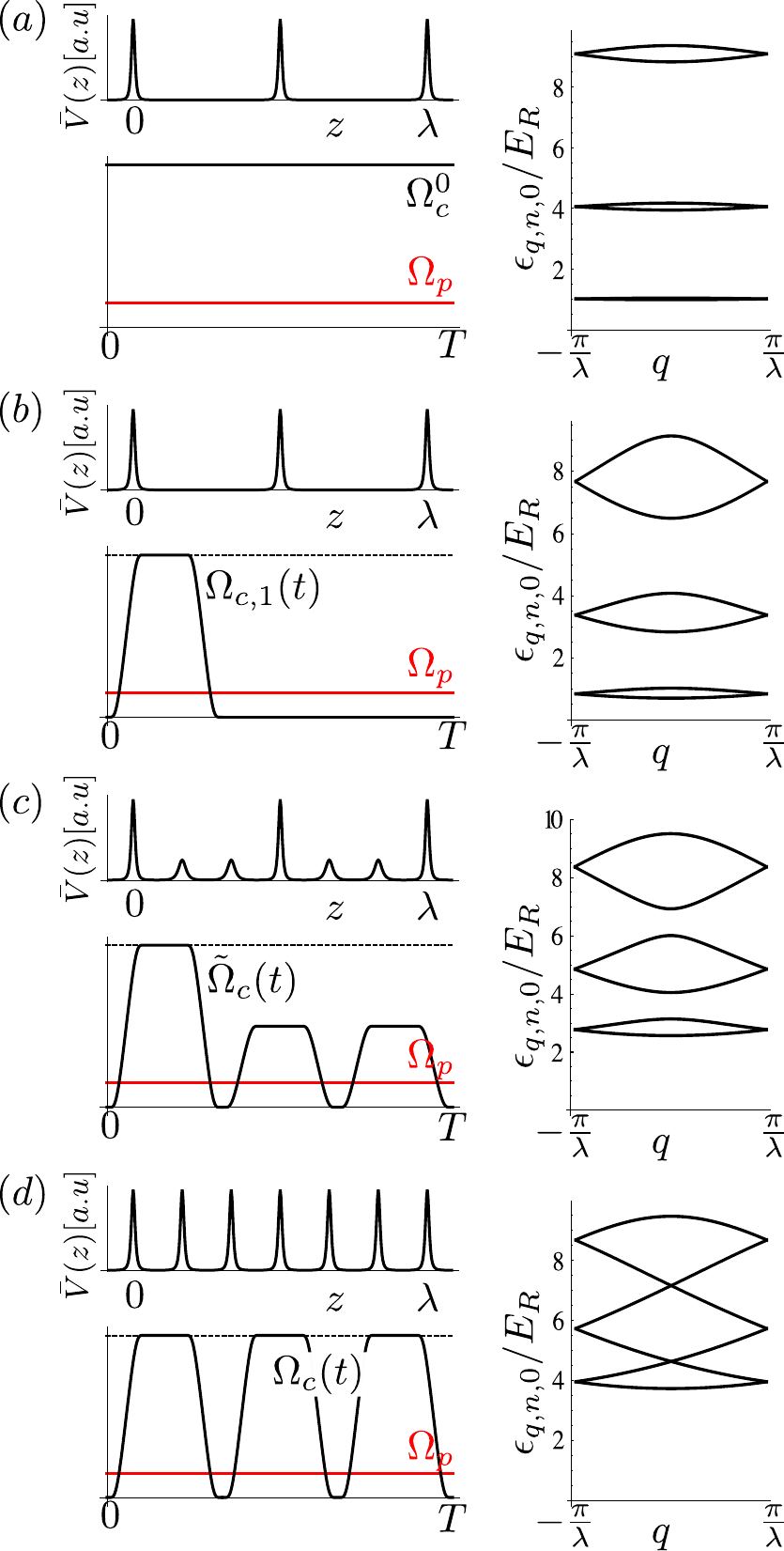}

\caption{Loading protocol for painted comb potential. Each panel shows the
time-averaged potential $\bar{V}(z)$ on the left upper part, the
time dependence of the Rabi frequencies $\Omega_{p}$ and $\Omega_{c}$
on the left lower part, and the lowest energy bands of $\bar{V}(z)$
on the right part. Panel (a) corresponds the end of the first step
when the auxiliary optical potential $V_{L}(z)$ is switched off and
the control Rabi frequency in the form of a standing wave $\Omega_{c}^{0}(z)=\Omega_{c}^{0}\sin(kz)$
with time-independent amplitude $\Omega_{c}^{(0)}$ is on. In panel
(b) we present the end of the second step when the time-dependent
amplitude $\Omega_{c,1}(t)$ of the control Rabi frequency generates
only the first flash in the sequence of flashes in the painting Protocol
III. Panels (c) and (d) show some intermediate and the final moments
of the step 3, respectively. During this step we gradually recover
the entire sequence of flashes of the painting protocol (see text).}

\label{fig:Fig6loading}
\end{figure}

We now describe protocols for loading atoms into the lowest band of
a painted potential. We consider a smooth and a comb potentials painted
with Protocol II and III, respectively, both with $M=3$. In both
cases we start with an atom described by the Hamiltonian (\ref{eq:threechannel})
with $\Omega_{c}=0$ but $\Omega_{p}\neq0$ in the presence of an
additional far-off-resonance optical potential
\begin{equation}
V_{L}(z)=V_{0}\cos^{2}(kz)\label{eq:VL}
\end{equation}
which is assumed to act equally on all three atomic states involved
in the $\Lambda$ scheme (in our calculations we use the value $V_{0}=10E_{R}$).
Initially, the atom is in the dark state $|D(z,t)\rangle=|g_{1}\rangle$
and its motional state is in the lowest Bloch band of the potential
$V_{L}(z)$. Note that $V_{\mathrm{na}}=0$ in this $\Lambda$ scheme
configuration. Another important thing to keep in mind is that the
probe Rabi frequency $\Omega_{p}$ remain constant during the entire
loading procedure, providing a nonzero gap between the dark and the
bright channels.

Let us first discuss a two-step loading into the smooth potential
$V_{W}(z)$, Eq.~(\ref{eq:sinpotential}), with $M=3$, $V=15E_{R}$,
and $W=3E_{R}$, painted with Protocol II, see Subsection~\ref{sec:Painting-smooth-potentials}.
The initial potential $V_{L}(z)$ and the corresponding three lowest
Bloch bands are shown in Fig.~\ref{fig:loadingsmooth}(a). In a first
step, we gradually increase the amplitude $\Omega_{c}$ of the ``moving''
control Rabi frequency $\Omega_{c}\sin(k[z-z_{0}(t)])$ with $z_{0}(t)=v_{0}t=\lambda t/T$
to the value $\Omega_{cW}$ such that the painted part of the potential
is equal to the offset $W$ of the desired potential $V_{W}(z)$.
The total (averaged) potential at this stage is $\bar{V}(z)=V_{L}(z)+W$
and the corresponding band structure are similar to those in Fig.~\ref{fig:loadingsmooth}(a)
but shifted upwards in energy by $W$. In the second step, we slowly
decrease $V_{0}$ to zero {[}say, as $V_{0}(t)=V_{0}\cdot(1-t/\tau)$
with $\tau\gg T${]} and increase $\Omega_{c}$ to its final value
$\Omega_{cV}=\Omega_{p}/\epsilon_{0}$ with $\epsilon_{0}=W+V/2$
{[}say, $\Omega_{c}(t)=\Omega_{cW}\cdot(1-t/\tau)+\Omega_{cV}\cdot t/\tau${]},
together with the corresponding adjustment of the velocity $dz_{0}(t)/dt$
from $v_{0}$ to $v(z)$ from Eq. (6). The time-averaged potential
and the corresponding low-energy band structure at some intermediate
time and at the end ($t=\tau)$ of this stage are shown in Figs.~\ref{fig:loadingsmooth}(b)
and \ref{fig:loadingsmooth}(c), respectively. We see that during
this loading protocol, the three lowest band of the initial potential
$V_{L}(z)$ merge together and form the lowest band of the final potential
(\ref{eq:sinpotential}), while all other bands are kept separated
by a finite energy gap, ensuring the transfer of the atom into the
final lowest band.

The loading procedure into the painted comb potential is similar and
has three steps. First, as in Ref.~\cite{Wang2017}, we slowly decrease
$V_{0}$ to zero and simultaneously ramp up the control Rabi frequency
to $\Omega_{c}^{0}(z)=\Omega_{c}^{0}\sin(kz)$ with $\Omega_{c}^{0}=\Omega_{p}/\epsilon_{0}$,
where $\epsilon_{0}$ determines the maximal value $E_{R}/\epsilon_{0}^{2}$
of the $V_{\mathrm{na}}(z,t)$ used for the painting, see Subsection\ \ref{sec:Painting-comb-potential}.
An atom which initially was in the lowest band of $V_{0}(z)$, is
now loaded into the lowest band of the time-independent potential
$V_{\mathrm{na}}(z)$, Eq.~(\ref{eq:Vsubwavelength}), with $z_{0}=0$,
see Fig.\ \ref{fig:Fig6loading}(a). During the next step we gradually
introduce a periodic time-dependence with period $T=2\pi/\omega$
of the control Rabi frequency, $\Omega_{c}^{0}\to\Omega_{c,1}(t)$,
where $\Omega_{c,1}(t)$ {[}see Fig.\ \ref{fig:Fig6loading}(b){]}
reproduces only the first flash in the painting protocol as discussed
in Subsection\ \ref{sec:Painting-comb-potential} and shown in Fig.\ \ref{fig:FigStroboscopic}.
At the end of this step, the time averaged potential is simply $V_{\mathrm{na}}(z)$
scaled down by a factor $\approx1/M$ ($\approx1/3$ in the considered
case) with the corresponding modification of the band structure. In
the final step we quickly switch on the motion of the potential peaks
$z_{0}(t)$ according to the Protocol III, see Fig.\ \ref{fig:FigStroboscopic}(c),
and then adiabatically increase the amplitude of the ``missing''
$M-1$ flashes in the painting protocol to their final value by changing
the time dependence of $\Omega_{c}(t)$ such that it acquires increasing
non-zero values during the times of missing flashes, $\Omega_{c,1}(t)\to\tilde{\Omega}_{c}(t)=\Omega_{c,1}(t)(1-t/\tau)+\Omega_{c}(t)t/\tau$
with $\tau\gg T$. {[}Note that switching on $z_{0}(t)$ at the beginning
of this step does not affect the potential $V_{\mathrm{na}}(z,t)$
because $z_{0}(t)\neq0$ when $\Omega_{c,1}(t)=0$.{]} The corresponding
time-averaged potentials $\bar{V}(z)$, time dependence of $\Omega_{c}(t)$,
and the low-energy band structures are presented in Figs.\ \ref{fig:Fig6loading}(c)
and \ref{fig:Fig6loading}(d) during and at the end of this step,
respectively. As in the case of the loading into the smooth potential,
the three lowest energy bands of $V_{0}(z)$ are merged into a single
band of the painted potential after the loading protocol, and all
higher bands always kept separated by a finite gap.

\section{Concluding remarks\label{sec:Conclusions}}

We have demonstrated that the subwavelength high optical potential
peaks created within an atomic $\Lambda$ system, can be used for
stroboscopic painting of optical potentials which changes on a subwavelength
scale, both in a smooth or in a sharp way. In contrast to usual stroboscopic
engineering of atomic Hamiltonians (see, for example, \cite{Goldman2014,nascimbene2015dynamic,Bukov2015}),
which becomes progressively accurate with increasing the stroboscopic
frequency $\omega$, the usage of the atomic $\Lambda$ scheme for
generating the subwavelength painting element introduces an upper
limit for the stroboscopic frequency, when bright states inevitable
present in the $\Lambda$ scheme become strongly coupled to the states
in the dark channel during stroboscopic dynamics. As a result, the
choice of $\omega$ is a compromise between the accuracy of painting
(as well as of the widths of avoided-crossings) and the involvement
of the bright channels, which limits the life-time of the system.
When drawing a smooth potential, the range of available stroboscopic
frequencies $\omega$ is generally broader than for the rapidly changing
potential because of a faster decay of high frequency Fourier components
of the time dependent drawing potential. In principle, the upper limit
on $\omega$ can be pushed to higher values by increasing the gap
(given by the control Rabi frequency $\Omega_{p}$) to the bright
states, but, to keep the same painting spatial resolution, this would
also require higher control Rabi frequencies $\Omega_{c}$ which could
be very challenging in practice. 

As a final remark, the proposed method can be used not only for painting
static potentials. One can also think of generating slowly time varying
potentials with subwavelength resolution, and using them for manipulations
of atomic systems. Examples of such manipulations could be splitting
of a potential well into two wells by raising a subwavelength barrier,
or merging two potential wells by decreasing the barrier between them.
As compared with the standard usage\textbf{ }(see, for example, \cite{Folling2007,Atala2013,Lohse2016})
of optical superlattices for such operations, our method provides
much more flexibility in spatial design of the potential barriers
and wells. 
\begin{acknowledgments}
We note that related and complementary work is being pursued by Subhankar
et. al \cite{Subhankar20192}.

M. L. is supported by the Polish National Science Centre project 2016/23/D/ST2/00721
and in part by PL-Grid Infrastructure. Work at Innsbruck is supported
by the ERC Synergy Grant UQUAM, by QTFLAG-QuantERA, and by the Institut
für Quanteninformation. 
\end{acknowledgments}

\appendix

\bibliographystyle{apsrev4-1}
\input{paper35_ML_MB.bbl}
\section{Born-Oppenheimer description of $\Lambda$ system}

\label{sec:details}

In the rotating wave approximation, the single-particle Hamiltonian
describing one-dimensional motion of an atom in the $\Lambda$ configuration
in Fig.~\ref{fig:Fig2atomicScheme} as discussed in the main text,
has the form:

\begin{eqnarray}
H & = & H_{\textrm{kin}}+H_{\textrm{\textrm{\ensuremath{\Lambda}}}}(z,t)=\label{eq:hamiltonian-1}\\
 & = & -\frac{\hbar^{2}}{2m_{a}}\partial_{z}^{2}+\hbar\left(\begin{array}{ccc}
0 & \Omega_{c}(z,t)/2 & 0\\
\Omega_{c}(z,t)/2 & -\Delta-i\Gamma/2 & \Omega_{p}/2\\
0 & \Omega_{p}/2 & 0
\end{array}\right),\label{eq:hamiltonian}
\end{eqnarray}
where $\Delta$ indicates the detuning of both Rabi frequencies $\Omega_{c}$
and $\Omega_{p}$ from the excited state $|e\rangle$with loss rate
$\Gamma$ due to spontaneous emission. In the Born-Oppenheimer (BO)
approximation, one neglects the kinetic energy such the right eigenvectors
of $H$ are determined by the atomic Hamiltonian $H_{\textrm{\ensuremath{\Lambda}}}$,
which are position and time dependent dark state \cite{Lacki2016}
(we prefer to use here different notations) 
\[
|D(z,t)\rangle=\frac{-\Omega_{p}|g_{1}\rangle+\Omega_{c}(z,t)|g_{2}\rangle}{E(z,t)}
\]
with zero eigenenergy, $E(z,t)=\sqrt{\Omega_{c}(z,t)^{2}+\Omega_{p}^{2}}$,
and two bright states 
\[
|B_{\pm}(z,t)\rangle=\frac{\Omega_{c}(z,t)|g_{1}\rangle+2E_{\pm}(z,t)|e\rangle+\Omega_{p}|g_{2}\rangle}{\mathscr{E}_{\pm}(z,t)},
\]
 where $\mathscr{E}_{\pm}(z,t)=\sqrt{E(z,t)^{2}+4E_{\pm}(z,t)^{2}}$
with local eigenenergies 
\begin{equation}
\hbar E_{\pm}(z,t)=\frac{\hbar}{2}\left[-\tilde{\Delta}\pm\sqrt{\tilde{\Delta}^{2}+E^{2}(z,t)}\right],\,\tilde{\Delta}=\Delta+i\Gamma/2.\label{eq:brightStateEnergies}
\end{equation}
The corresponding left eigenstates of $H_{\Lambda}$, which form a
biorthogonal system with the above right eigenstates, $\langle\alpha(z,t)|\beta(z,t)\rangle=\delta_{\alpha\beta}$,
are
\[
\langle D(z,t)|=\frac{-\Omega_{p}\langle g_{1}|+\Omega_{c}(z,t)\langle g_{2}|}{E(z,t)},
\]
\[
\langle B_{\pm}(z,t)|=\frac{\Omega_{c}(z,t)\langle g_{1}|+2E_{\pm}(z,t)\langle e|+\Omega_{p}\langle g_{2}|}{\mathscr{E}_{\pm}(z,t)},
\]
The BO local eigenbasis $\{|D(z,t)\rangle,|B_{+}(z,t)\rangle,|B_{-}(z,t)\rangle\}\equiv\{|\alpha(z,t)\rangle\}$
can be used for expansion of an arbitrary atomic state, $|\psi(z,t)\rangle=\sum_{\alpha=D,B_{\pm}}\psi^{(\alpha)}(z,t)|\alpha(z,t)\rangle$,
and the Hamiltonian for the wave functions $\{\psi^{(\alpha)}(z,t)\}$,
obtained from the Schrödinger equation $i\hbar\partial_{t}|\psi(z,t)\rangle=H|\psi(z,t)\rangle$
with the Hamiltonian $H$ from Eq.~(\ref{eq:hamiltonian}), reads

\begin{equation}
H_{BO}(z,t)=-\frac{\hbar^{2}}{2m}[\partial_{z}+\hat{A}(z,t)]^{2}+\hbar\hat{E}(z,t)-i\hbar\hat{A}_{t}(z,t),\label{eq:BOham}
\end{equation}
where the matrix $\hat{A}(z,t)$ has matrix elements $\hat{A}_{\alpha\beta}(z,t)=\langle\alpha(z,t)|\partial_{z}|\beta(z,t)\rangle$,
$\hat{E}(z,t)$ is the diagonal matrix of BO local eigenenergies,
$\hat{E}(z,t)=\textrm{diag}[0,E_{+}(z,t),E_{-}(z,t)]$, and The matrix
elements of $\hat{A}_{t}(z,t)$ are $\hat{A}_{t,\alpha\beta}(z,t)=\langle\alpha(z,t)|\partial_{t}|\beta(z,t)\rangle$.
Explicitly, the matrices $\hat{A}(z,t)$ and $\hat{A}_{t}(z,t)$ are
given by

\[
\hat{A}(z,t)=\frac{\Omega_{p}\partial_{z}\Omega_{c}(z,t)}{E(z,t)}\left(\begin{array}{ccc}
0 & -\mathscr{E}_{+}^{-1}(z,t) & -\mathscr{E}_{-}^{-1}(z,t)\\
\mathscr{E}_{+}^{-1}(z,t) & 0 & D(z,t)\\
\mathscr{E}_{-}^{-1}(z,t) & -D(z,t) & 0
\end{array}\right)
\]
and

\[
\hat{A}_{t}(z,t)=\frac{\Omega_{p}\partial_{t}\Omega_{c}(z,t)}{E(z,t)}\left(\begin{array}{ccc}
0 & -\mathscr{E}_{+}^{-1}(z,t) & -\mathscr{E}_{-}^{-1}(z,t)\\
\mathscr{E}_{+}^{-1}(z,t) & 0 & D(z,t)\\
\mathscr{E}_{-}^{-1}(z,t) & -D(z,t) & 0
\end{array}\right)
\]
with 
\[
D(z,t)=\frac{\Omega_{c}(z,t)}{2\Omega_{p}}\frac{\tilde{\Delta}}{\tilde{\Delta}^{2}+E^{2}(z,t)}
\]

The expressions for $\hat{A}(z,t)$ and $\hat{A}_{t}(z,t)$ become
substantially simpler for $\tilde{\Delta}=0$,

\[
\hat{A}(z,t)=\frac{\Omega_{p}\partial_{z}\Omega_{c}(z,t)}{\sqrt{2}E(z,t)^{2}}\left(\begin{array}{ccc}
0 & -1 & -1\\
1 & 0 & 0\\
1 & 0 & 0
\end{array}\right)
\]

and

\[
\hat{A}_{t}(z,t)=\frac{\Omega_{p}\partial_{t}\Omega_{c}(z,t)}{\sqrt{2}E(z,t)^{2}}\left(\begin{array}{ccc}
0 & -1 & -1\\
1 & 0 & 0\\
1 & 0 & 0
\end{array}\right).
\]

In the expansion of the kinetic energy term in Eq.~(\ref{eq:BOham})
in powers of $\hat{A}$, the off-diagonal elements of $\hat{A}$ and
$\hat{A}^{2}$, together with $\hat{A}_{t}$, define the nonadiabatic
couplings between the dark and the bright states, while the diagonal
elements in $\hat{A}^{2}$ give rise to nonadiabatic potentials in
the BO channels. For the dark state this nonadiabatic potential is
\[
V_{\mathrm{na}}(z,t)=\frac{\hbar^{2}}{2m}\left\{ \frac{\Omega_{p}\partial_{z}\Omega_{c}(z,t)}{\Omega_{p}^{2}+[\partial_{z}\Omega_{c}(z,t)]^{2}}\right\} ^{2}
\]
 and takes the form of Eq.~(\ref{eq:Vsubwavelength}) in the main
text with $\epsilon(t)=\Omega_{p}/\Omega_{c}(t)$ and $z_{0}(t)$
for $\Omega_{c}(z,t)=\Omega_{c}(t)\sin[k(z-z_{0}(t)]$. The effects
of the bright states on the atomic dynamics in the dark state due
to the nonadiabatic couplings can be decreased by making larger gaps
$\Delta E_{B\pm}=\min_{z}\left|\hbar E_{\pm}(z)\right|$ between the
BO channels, which is achieved by increasing the values of the Rabi
frequencies. In this case, the dynamics of an atom in the dark state
is dominated by the time-averaged nonadiabatic potential $\bar{V}(z)$
from Eq.~(\ref{eq:runningapproximation}).

\section{Time dependence of $\Omega_{c}$ for Protocol III.}

\label{sec:AppTimeDependence}

Here we specify the time-dependence of the amplitude $\Omega_{c}(t)$
of the control Rabi frequency we use in numerical simulation for the
Protocol III, which generates the time evolution of the height of
the potential peaks shown in Fig.~\ref{fig:FigStroboscopic}c. Following
this figure, the amplitude $\Omega_{c}(t)$ should be zero at times
$t=0,T/M,2t/M,\ldots$, and in between these times it increases to
its maximal value $\Omega_{c}^{\textrm{max.}}$ during time $\tau_{s}$,
stays constant for time $\tau_{h}$, and decreases back to zero during
time $\tau_{s}$. For the elementary period $t=[0,T/M${]}, we define
$\Omega_{c}(t)$ as $\Omega_{c}(t)=\Omega_{c}^{\textrm{max.}}f(t)$
with the function $f(t)$ given by 

\begin{equation}
f(t)=\begin{cases}
0, & t\in[0,\tau_{h}/2),\\
h(t), & t\in[\tau_{h}/2,\tau_{h}/2+\tau_{s}),\\
1, & t\in[\tau_{h}/2+\tau_{s},3\tau_{h}/2+\tau_{s}),\\
h(3\tau_{h}/2+2\tau_{s}-t), & t\in[3\tau_{h}/2+\tau_{s},3\tau_{h}/2+2\tau_{s}),\\
0, & t\in[3\tau_{h}/2+2\tau_{s},2\tau_{h}+2\tau_{s}).
\end{cases}\label{eq:ef}
\end{equation}
where
\begin{eqnarray*}
h(x) & = & \begin{cases}
0, & x\leq0\\
\exp(-1/x)e^{2}/2, & x<1/2\\
1-\exp[-1/(1-x)]e^{2}/2, & 1/2<x<1\\
1, & x\geq1.
\end{cases}
\end{eqnarray*}
This choice avoids discontinuities of $\Omega_{c}(t)$ and its derivatives
which lead to slow decay of the matrix elements coupling different
Floquet blocks (as discussed in Section \ref{subsec:3DrivingWithBrightState}).
\end{document}

%% file: paper35_ML_MB.bbl
%